\documentclass[
amsmath, %
floatfix, %
twocolumn, %
reprint, %
prb, %
aps, %
]{revtex4-2}

\renewcommand{\thispagestyle}[1]{}
\usepackage[T1]{fontenc}
\usepackage[utf8]{inputenc}
\usepackage{graphicx}
\usepackage{latexsym}
\usepackage{amsthm}
\usepackage{color}
\usepackage{mathtools}
\usepackage{xcolor}
\usepackage[section]{placeins}

\newcommand{\mfrac}[2]{\tfrac{#1}{#2}}

\usepackage[]{newtxtext}
\usepackage[nosymbolsc,smallerops,bigdelims]{newtxmath}
\DeclareMathAlphabet{\mathcal}{OMS}{cmsy}{m}{n}
\renewcommand{\mfrac}[2]{#1/#2}

\DeclareMathAlphabet{\mathcalb}{OMS}{cmsy}{b}{n}
\usepackage{bm}
\usepackage{hyperref}
\hypersetup{
	colorlinks,
	linkcolor={blue!90!black},
	citecolor={blue!90!black},
	urlcolor={blue!90!black}
}
\makeatletter
\renewcommand*{\eqref}[1]{%
	\hyperref[{#1}]{\textup{\tagform@{\ref*{#1}}}}%
}
\makeatother

\DeclarePairedDelimiter\lr{\lparen}{\rparen}
\DeclarePairedDelimiter\Lr{\lbrack}{\rbrack}
\DeclarePairedDelimiter\LR{\lbrace}{\rbrace}
\DeclarePairedDelimiter\abs{\lvert}{\rvert}
\DeclarePairedDelimiter\avg{\langle}{\rangle}
\DeclarePairedDelimiter\norm{\lvert}{\rvert}

\DeclarePairedDelimiterX{\comm}[2]{\lbrack}{\rbrack}{#1, #2}
\DeclarePairedDelimiterX{\acomm}[2]{\lbrace}{\rbrace}{#1, #2}

\DeclarePairedDelimiter\ket{\lvert}{\rangle}

\DeclarePairedDelimiterX{\braket}[2]{\langle}{\rangle}{#1\delimsize\vert #2}
\DeclarePairedDelimiterX{\ketbra}[2]{\rvert}{\lvert}{#1 \delimsize\rangle\!\delimsize\langle #2}
\DeclarePairedDelimiterX{\matrixel}[3]{\langle}{\rangle}{#1 \delimsize\rvert #2 \delimsize\lvert #3}

\renewcommand{\tilde}[1]{\widetilde{#1}}

\newcommand{\e}{\mathrm{e}}

\newcommand{\DxDy}{\iiint\limits_{-\infty}^{\infty}\!\!\mathrm{D}x_P\mathrm{D}x_Q\mathrm{D}x_B}
\newcommand{\Dxxx}{\int\!\!\!\mathrm{D}\bm{x}}
\newcommand{\lkaka}{\lambda_{k\alpha k'\alpha'}}

\newcommand{\nka}{\nu_{k\alpha}}
\newcommand{\Heff}{H_{\mathrm{eff}}}
\newcommand{\pka}{p_{\alpha}^{(k)} }
\newcommand{\qka}{q_{\alpha}^{(k)} }
\newcommand{\pkka}{p_{\alpha}^{(k')}}
\newcommand{\qkka}{q_{\alpha}^{(k')}}
\newcommand{\Mpq}{\mathcal{M}_{pq}}
\newcommand{\Trpq}{\mathrm{Tr}_{pq}}
\newcommand{\Rkk}{\mathcal{R}_{kk'}}
\newcommand{\Ukk}{\mathcal{U}_{kk'}}

\newcommand{\Qaa}{\mathcal{Q}_{\alpha\alpha'}}

\newcommand{\JOBM}{\frac{J_0\beta}{M}}
\newcommand{\JBM}{\frac{J\beta}{M}}

\newcommand{\pkia}{p_{i\alpha}^{(k)}}
\newcommand{\pkja}{p_{j\alpha}^{(k)}}
\newcommand{\qkia}{q_{i\alpha}^{(k)}}
\newcommand{\qkja}{q_{j\alpha}^{(k)}}
\newcommand{\pkkaa}{p_{\alpha'}^{(k')}}
\newcommand{\qkkaa}{q_{\alpha'}^{(k')}}
\newcommand{\JBMa}[2][]{\frac{#1 J\beta}{#2 M}}
\newcommand{\eBH}{\exp(-\beta\Heff)}
\newcommand{\pkkiaa}{p_{i\alpha'}^{(k')}}
\newcommand{\qkkiaa}{q_{i\alpha'}^{(k')}}
\newcommand{\Pkaka}{\prod_{k\alpha k'\alpha'}}
\newcommand{\Skaka}{\sum_{k\alpha k'\alpha'}}
\newcommand{\Pka}{\prod_{k\alpha}}
\newcommand{\Ska}{\sum_{k\alpha}}
\newcommand{\Si}{\sum_i}
\newcommand{\SP}{\tilde{P}}
\newcommand{\SQ}{\tilde{Q}}
\newcommand{\SJ}{j}
\newcommand{\intid}{\int\limits_{-\infty}^{\infty}\!\!\mathrm{d}}

\begin{document}
\title{Emergence of a superglass phase
in the random hopping Bose-Hubbard model}
\author{Anna M. Piekarska}
\email{a.piekarska@intibs.pl}
\affiliation{Institute of Low Temperature and Structure Research, Polish Academy of Sciences, 
Ok\'{o}lna 2, 50-422 Wroc\l{}aw, Poland}
\author{Tadeusz K. Kope\'{c}}
\affiliation{Institute of Low Temperature and Structure Research, Polish Academy of Sciences, 
Ok\'{o}lna 2, 50-422 Wroc\l{}aw, Poland}
\begin{abstract}
We study an experimentally feasible system of strongly correlated bosons with random hoppings,
described by the infinite-range Bose-Hubbard model on a lattice
with hopping integrals given by independent random variables of Gaussian distribution with non-zero mean.
We solve this quantum model in the thermodynamic limit, employing the replica method and the Trotter-Suzuki formula.
We find and describe a superglass phase that emerges at the interface between glass and superfluid phases.
Both glassy and long-range orderings are present in the superglass and compete with each other, as revealed by the anticorrelation of their order parameters.
We present phase diagrams in various cross-sections of the multidimensional space of system parameters.
In selected parameter subspaces, we compare the results to those of non-disordered, diagonally-disordered, and once celebrated spin-glass systems.
\end{abstract}
\maketitle
%
\section{Introduction}
The interplay between interactions and disorder in quantum many-body systems is a field
where a lot remains to be explored~\cite{Zapf2014_RoMP86}.
Its complexity and relevance stem from the fact that these two effects
have opposing impacts on particle localization~\cite{Giamarchi1988_PRB37,Schreiber2015_S349},
and thus their competition leads to interesting physical phenomena.
Moreover, the interest in this kind of systems increases
due to the rapidly developing quantum simulation methods~\cite{Feynman1982_IJoTP21,Georgescu2014_RoMP86},
such as, e.g., the creation of optical lattices~\cite{Jaksch2005_AoP315}.
They allow for an experimental realization of the theoretically studied systems~\cite{Greiner2002_N415,Morrison2008_NJoP10},
and this provides a two-way correspondence in the study of model Hamiltonians.

Disordered systems can be classified based on the character of the disorder.
In bosonic lattice systems, the most commonly studied type is \emph{diagonal}
disorder~\cite{Fisher1989_PRB40,Singh1992_PRB46,Weichman2008_PRB77,Buonsante2009_PRA79,Krueger2009_PRB80,
Gurarie2009_PRB80,Bissbort2010_PRA81,Niederle2013_NJoP15,Lin2017_SR7},
i.e., the randomness in the system is present in the chemical potential.
It was found that in such a case there is no direct transition between superfluid and Mott insulator phases,
as a Bose glass phase emerges between them upon introduction of disorder~\cite{Pollet2009_PRL103}.
The case discussed in this work is the less explored one of random interactions
\cite{Prokofev2004_PRL92,Sengupta2007_PRL99,Buonsante2007_LP17,Bissbort2010_PRA81,Piekarska2018_PRL120,Piekarska2020_JoSMTaE2020},
called the \emph{off-diagonal} disorder.
Such a system is frustrated and thus the glassy phase that emerges in it differs from the Bose glass \cite{Yu2012_PRB85}.
Moreover, describing the Bose glass in terms of the Edwards-Anderson order parameter \cite{Edwards1975_JoPFMP5}
requires defining it via particle density fluctuations \cite{Thomson2014_EL108}.
Here, we aim to perform an analysis of the phase diagram of the off-diagonal case.

In magnetic systems, the off-diagonal kind of disorder
is a vital ingredient of once very popular spin-glass systems~\cite{Edwards1975_JoPFMP5,Sherrington1975_PRL35},
for which the replica symmetry breaking phenomenon has been found first~\cite{Parisi1979_PRL43}.
An essential feature of disordered interaction is \emph{frustration}.
There are many nearly degenerate local minima of energy in a frustrated system,
separated by energy barriers of significant height~\cite{Binder1986_RoMP58}.
Thus, the system may remain in an excited state for a very long time,
depending on the history of its evolution~\cite{Palmer1982_AiP31}.
Due to this, we expect the off-diagonal disorder case
to be significantly different from the diagonal one.
The possibility of quantum tunneling enables transitions
between local energy minima without the need for thermal fluctuations,
which effectively softens this slow relaxation effect~\cite{Wu1991_PRL67}.
This property was shown to make quantum spin glasses significantly different
from their classical counterparts~\cite{Bray1980_JoPCSSP13}.
In particular, a quantum phase transition is present in such systems~\cite{Wu1993_PRL71},
which has also been addressed theoretically~\cite{Goldschmidt1990_PRL64,Usadel1986_SSC58,Guo1994_PRL72,Rieger1994_PRL72}.

The other component of rich physics of disordered many-body systems is strong correlations 
that lead to various forms of collective behavior, like
superfluidity~\cite{Kapitza1938_N141,Allen1938_N142}
or high-$T_\mathrm{c}$ superconductivity~\cite{Bednorz1986_ZfPBCM64}.
Superfluidity can be intrinsically found, e.g., in liquid helium~\cite{Kapitza1938_N141,Allen1938_N142},
but in recent years, Bose-Einstein condensation in ultracold dilute gases~\cite{Anderson1995_S269}
has emerged as a framework offering easier access to investigate its properties~\cite{Bloch2008_RoMP80}.
Introducing periodic potentials to such systems expanded their usefulness to quantum simulation.
The availability of such frameworks opened a new way to study disordered systems~\cite{Morrison2008_NJoP10,Ahufinger2005_PRA72}.

Competition between glassiness and long-range order has been known,
for example, in spin glasses,
where ferromagnetism was found to destroy the glass ordering~\cite{Binder1986_RoMP58}.
However, solid ${}^{4}$He was found to exhibit a superglass phase~\cite{Hunt2009_S324}
(initially classified as a supersolid~\cite{Kim2004_N427}),
in which the glass and superfluid orders coexist.
Several theoretical works~\cite{Carleo2009_PRL103,Tam2010_PRL104,Yu2012_PRB85,Angelone2016_PRL116}
emerged to confirm the existence of this new phase and describe it adequately.
It was shown~\cite{Yu2012_PRB85} that the two orders compete within the superglass phase.
Nevertheless, they can indeed be present alongside each other.

In this paper, we consider a system of strongly correlated bosons with normal-distributed random hopping of nonzero mean.
The model is fully connected, however, experimental realization with optical lattices is possible~\cite{Piekarska2018_PRL120}.
We study the competition between the glassy and superfluid orders,
which leads to the emergence of the superglass (SG) phase,
apart from the usual ordered phases: glass (GL) and superfluid (SF),
as well as the disordered (DI) one.
Examining the behavior of the order parameters in the studied phases,
we find the anticorrelation in agreement with Ref.~\cite{Yu2012_PRB85}.
We obtain the phase diagrams based on conditions following from the Landau theory
and the stability criterion for the replica-symmetric solution ~\cite{Piekarska2022a}.
In particular, we find the SG phase as the part of the superfluid region with broken replica symmetry.
We evaluate these conditions by solving numerically self-consistent equations
that arise after a derivation that follows a similar scheme as our previous work~\cite{Piekarska2018_PRL120}
and its spin-glass and quantum-spin-glass predecessors~\cite{Sherrington1975_PRL35,Usadel1988_NPBPS5}.
First, we use the replica method~\cite{Sherrington1975_PRL35} and the Trotter-Suzuki formula~\cite{Suzuki1976_CiMP51}
to map the problem onto an effective classical model,
to which we then apply the saddle-point method in the thermodynamic limit.
The saddle point solution gives us the desired self-consistent equations.
Importantly, in the method used by us, the averaging over disorder is done exactly,
as an analytical integration over the entire distribution, as opposed to averaging several realizations of disorder.

The text is organized as follows.
First, in Section~\ref{sec:methods}, we present a brief description of the conducted analytical derivation
and the subsequent numerical treatment of the obtained equations.
Next, in Section~\ref{sec:crit}, we establish the conditions from critical lines
and predict some of their behavior.
Then, in Section~\ref{sec:results}, we present and analyze the numerically calculated phase diagrams.
Finally, in Section~\ref{sec:summary}, we comment on the obtained results, relate the current work to existing knowledge,
and suggest possible directions of future research in this field.
Outside the main text, in Appendix~\ref{app:compare},
we show a quantitative comparison of the limiting case of our results with previous work,
while in Appendices~\ref{app:limits}--\ref{app:Delta},
we present more detailed derivation steps mentioned in Sections~\ref{sec:methods} and \ref{sec:crit}.
\section{Model and methods}\label{sec:methods}
\subsection{Model}
The Bose-Hubbard Hamiltonian for the system of $N$ interacting bosons reads
\begin{equation}\begin{split}
	H =& -\sum_{i<j}J_{ij}\lr*{a_i^\dagger a_j+a_j^\dagger a_i}\\
	   & + \frac{U}{2}\Si \hat{n}_i\lr*{\hat{n}_i-1} - \mu \Si \hat{n}_i,
\end{split}\end{equation}
where $a_i$ ($a_i^\dagger$) are the annihilation (creation) operators for the site $i$
and $\hat{n}_i=a_i^\dagger a_i$ are the particle number operators,
while $\mu$ and $U$ denote the chemical potential and on-site interaction strength, respectively.
$J_{ij}$ are independent random variables describing the hopping integrals between sites $i$ and $j$.
They are given by Gaussian distribution with the mean $J_0/N$ and variance $J^2/N$,
following Ref.~\cite{Sherrington1975_PRL35}.
As the disorder in the studied system is \emph{quenched},
we need to average calculated quantities over its distribution, i.e.,
\begin{equation}
	\Lr*{X}_J = \prod_{i<j}\Lr*{\intid J_{ij} \frac{\exp\lr*{-\frac{\lr*{J_{ij}-J_0/N}^2}{2J^2/N}}}{J\sqrt{2\pi/N}}}X,
\end{equation}
where by $\Lr*{\cdots}_J$, we denote the desired average over the disorder,
while $X$ is the averaged quantity, dependent on all $J_{ij}$.

To make the calculations more convenient,
we transform the Hamiltonian to the basis of quasi-momentum and quasi-position, i.e.,
\begin{equation}
	\hat{P} = \frac{i}{\sqrt{2}}\lr*{a^\dagger-a}, \qquad\hat{Q} = \frac{1}{\sqrt{2}}\lr*{a^\dagger+a}.
\end{equation}
The transformed Hamiltonian reads
\begin{equation}
	H = -\sum_{i < j} J_{ij}\lr*{\hat{P}_i\hat{P}_j +\hat{Q}_i\hat{Q}_j}
	+ \Si\lr*{\tilde{U}\hat{n}_i^2 - \tilde{\mu}\hat{n}_i},
\end{equation}
where we have introduced $\tilde{\mu}= \mu+\frac{U}{2}$ and $\tilde{U}= \frac{U}{2}$.

\subsection{Order parameters}
The natural order parameter in a strongly correlated bosonic system
is the \emph{superfluid order parameter} \cite{Fisher1989_PRB40}
\begin{equation}\label{eq:defD}
	\varDelta = \avg{a_i},
\end{equation}
where $\avg{\cdot} = \mathrm{Tr} \cdot \mathrm{e}^{-\beta H}\!/\,\mathrm{Tr} \, \mathrm{e}^{-\beta H}$
denotes the thermodynamic average.
However, it does not capture the glass ordering, i.e., the quenched disorder,
as the locally frozen phases of complex wave functions average to zero over the whole material,
despite their uniquely determined values.
This behavior can in turn be identified based on the \emph{Edwards-Anderson order parameter} \cite{Edwards1975_JoPFMP5}
\begin{equation}\label{eq:defq}
	\mathcal{Q}_{\mathrm{EA}} = \frac{1}{N}\sum_i \Lr*{\norm*{\avg{a_i}}^2}_J.
\end{equation}
Thus, combining both $\varDelta$ and $\mathcal{Q}_{\mathrm{EA}}$, we are able to identify the
\emph{disordered} phase (also called \emph{Mott insulator}; characterized with $\mathcal{Q}_{\mathrm{EA}}=0$, $\varDelta=0$),
the \emph{glassy} phase ($\mathcal{Q}_{\mathrm{EA}}> 0$, $\varDelta=0$) and the \emph{superfluid} phase ($\mathcal{Q}_{\mathrm{EA}}> 0$, $\varDelta> 0$).
A phase with $\mathcal{Q}_{\mathrm{EA}}=0$ and $\varDelta> 0$ is forbidden by symmetries in the system, as we show further.

\subsection{Effective classical model}
To obtain the free energy $F$ averaged over the $J_{ij}$ distributions,
we employ the replica trick \cite{Sherrington1975_PRL35},
\begin{equation}
	F = -\frac{1}{\beta} \Lr*{\ln Z}_J = - \lim_{n\rightarrow 0}\frac{1}{\beta n}\lr[\Big]{\Lr*{Z^n}_J-1},
\end{equation}
where $\beta=1/(k_{\mathrm{B}}T)$ (we put $k_{\mathrm{B}}=1$ throughout the paper),
$Z = \mathrm{Tr} \exp(-\beta H)$ is the partition function, and we introduce the replicated Hamiltonian $H_{\mathrm{repl}}$,
\begin{equation}
	Z^n = \mathrm{Tr} \exp \lr*{-\beta\sum_{\alpha=1}^n H_\alpha} \equiv \mathrm{Tr} \exp \lr*{-\beta H_{\mathrm{repl}}},
\end{equation}
in which each $H_\alpha$ is a copy of the original Hamiltonian.

Since various terms of the Hamiltonian do not commute, we apply the Trotter-Suzuki formula,
\begin{multline}
	\exp\lr*{-\beta H_{\mathrm{repl}}} =
		\lim_{M\rightarrow\infty} \Lr[\Bigg]{
			\exp\lr*{-\frac{\beta }{M}H_P}\exp\lr*{-\frac{\beta}{2M}H_n}\\
			\times\exp\lr*{-\frac{\beta}{M}H_Q}\exp\lr*{-\frac{\beta}{2M}H_n}
		}^M,
\end{multline}
where we have split the replicated Hamiltonian
\begin{equation}
	H_{\mathrm{repl}}=H_P+\frac{1}{2}H_n+H_Q+\frac{1}{2}H_n,
\end{equation}
introducing the following terms:
\begin{subequations}
\begin{align}
	H_P &= -\sum_\alpha\sum_{i < j} J_{ij}\hat{P}_{i\alpha}\hat{P}_{j\alpha},\\
	H_Q &= -\sum_\alpha\sum_{i < j} J_{ij}\hat{Q}_{i\alpha}\hat{Q}_{j\alpha},\\
	H_n &= \tilde{U}\sum_{i\alpha}\hat{n}_{i\alpha}^2-\tilde{\mu}\sum_{i\alpha}\hat{n}_{i\alpha}.
\end{align}
\end{subequations}
Next, between each pair of consecutive exponents,
we insert a summation over a complete set of eigenvectors of either $\hat{P}$ or $\hat{Q}$
(i.e., $\hat{P}\ket{p}=p\ket{p}$, etc.),
such that the matrix elements of the $H_P$ and $H_Q$ terms can be evaluated.
The resulting expression reads
\begin{align}
	\!Z^n = \mathrm{Tr}_{pp'qq'}\prod_{k=1}^M&
		\exp\Lr*{
			\frac{\beta}{M}\sum_\alpha\sum_{i < j}
				J_{ij}\lr*{\pkia\pkja+\qkia\qkja}
		} \nonumber \\
	&\times \braket*{p^{(k)}}{p'^{(k)}}\matrixel*{p'^{(k)}}{e^{-\frac{\beta H_n}{2M}}}{q^{(k)}}\\ 
	&\times \braket*{q^{(k)}}{q'^{(k)}}\matrixel*{q'^{(k)}}{e^{-\frac{\beta H_n}{2M}}}{p^{(k+1)}}, \nonumber
\end{align}
where the trace $\mathrm{Tr}_{pp'qq'}$ means
\begin{align}
	\MoveEqLeft \mathrm{Tr}_{pp'\!qq'} (\cdot) \nonumber \\[3pt]
		\!&=\sum_{p_{11}^{(1)}}\ldots\!\!\!\sum_{p_{Nn}^{(M)}}
		\!\sum_{p_{11}'^{(1)}}\!\ldots\!\!\!\sum_{p_{Nn}'^{(M)}}
		\!\sum_{q_{11}^{(1)}}\!\ldots\!\!\sum_{q_{Nn}^{(M)}}
		\!\sum_{q_{11}'^{(1)}}\!\ldots\!\!\!\sum_{q_{Nn}'^{(M)}}\! (\cdot),
\end{align}
and we define $\ket{p^{(k)}} \equiv \bigotimes_{i,\alpha}\ket{\pkia}$, etc.

As the expression can be now factorized into parts depending on one $J_{ij}$ only,
each of these parts may be averaged separately.
Therefore, we perform the averaging over Gaussian distributions,
resulting in the following expression:
\begin{equation}\begin{split}\label{eq:averaged}
	\Lr*{Z^n}_{J} = {}&\Trpq \mathcal{M}_{pq}\\
	&\times \prod_{i < j} \exp \LR[\Bigg]{
		\frac{J^2\beta^2}{2M^2N}\Lr*{
			\Ska\lr*{\pkia\pkja + \qkia\qkja}
		}^2\\
		&\mskip80mu+\frac{\beta J_0}{MN} \Ska\lr*{
			\pkia\pkja + \qkia\qkja
		}
	},
\end{split}\end{equation}
where $\mathcal{M}_{pq}$ is a product of matrix elements,
\begin{equation}\begin{split}
	\mathcal{M}_{pq} &= \prod_i \mathcal{M}_{pq}^{(i)}\\
	&=\prod_{ik\alpha}
		\matrixel*{\pkia}{e^{-\frac{\beta H_n}{2M}}}{\qkia}
		\matrixel*{\qkia}{e^{-\frac{\beta H_n}{2M}}}{p_{i\alpha}^{(k+1)}}.
\end{split}\end{equation}
At this point, the partition function has neither quantum nor random components.
The last trace of the quantum nature of the problem lies in the $\Mpq$ factor,
which, however, can be seen as an ad-hoc-defined function of $\LR[\big]{\pkia}$ and $\LR[\big]{\qkia}$.
Therefore, we have mapped the original problem onto an effective classical one,
at the cost of adding an additional time-like dimension,
introduced in the Trotter step.

\subsection{Self-consistent equations}
To take the thermodynamic limit, we need all the terms in $Z^n$ to involve a single site only.
All the site-mixing terms can be expressed in the form $(\sum_{i}T_{i})^{2}$,
to which we apply the Hubbard-Stratonovich transformation,
introducing a new variable that couples to the now single-site term $\sum_{i}T_{i}$.
In the thermodynamic limit, we use the saddle point method to obtain the effective free energy.
Finally, we take the limit of $n\rightarrow 0$ and arrive at
\begin{multline}\label{eq:FreeEnergy}
	\!\!\!\mathcal{F} =
		2\lr[\Big]{\JBM}^{2}\sum_{kk'}(\Rkk^2+\Ukk^2) + J_{0}\beta\varDelta^2 - 2\lr[\Big]{\frac{J\beta}{M}}^{2}(q^2+u^2)\\
		- \DxDy \ln \Trpq \exp \lr*{-\beta \mathcal{H}},
\end{multline}
where we use the notation $\mathrm{D}x \equiv \mathrm{d}x\exp(-x^2)$ for Gaussian integrals.
The details of this derivation can be found in Appendix~\ref{app:limits}.
The corresponding effective Hamiltonian is
\begin{multline}\label{eq:effham}
	\!\!\!\!\!\!-\beta\mathcal{H} =
		\varDelta\JOBM \sum_{k}\lr*{p_{k}+q_{k}} + \ln \mathcal{M}_{pq}\\
		+\frac{2J\beta}{M}\sum_{k}\lr*{ x_{B}\sqrt{\frac{u}{2}}(p_{k}+q_{k}) + \sqrt{\frac{q-u}{2}}(x_{P}p_{k}+x_{Q}q_{k}) }\\
		+2\lr[\Big]{\JBM}^{2}\sum_{kk'}\Lr[\Big]{
			(\Rkk-q)\lr*{p_{k}p_{k'} + q_{k}q_{k'}}\\
			+2(\Ukk-u)p_{k}q_{k'}
		}.
\end{multline}
Here, $\Rkk$, $\Ukk$, $\varDelta$, $q$, and $u$ are the variables introduced in the saddle point method,
and defined self-consistently as
\begin{subequations}
\begin{eqnarray}
	\label{eq:selfcon_R} \Rkk &=& \DxDy \avg{p_kp_{k'}},\\
	\Ukk &=& \DxDy \avg{p_kq_{k'}},\\
	\label{eq:selfcon_D} \varDelta &=& \DxDy \avg{p_k},\\
	\label{eq:selfcon_q} q &=& \DxDy \avg{p_k}^2\\
	\label{eq:selfcon_u} u &=& \DxDy \avg{p_k}\avg{q_k}.
\end{eqnarray}
\end{subequations}
The averages are thermal averages taken with the effective Hamiltonian \eqref{eq:effham}, i.e.,
\begin{equation}
	\avg{A} = \frac{\Trpq A \exp(-\beta\mathcal{H})}{\Trpq \exp(-\beta\mathcal{H})},
\end{equation}
which makes Eqs.~(\ref{eq:effham}) and (\ref{eq:selfcon_R})--(\ref{eq:selfcon_u}) self-consistent.
The new variables originate from decoupling of second-order ($\varDelta$) and fourth-order ($\Rkk$, $\Ukk$, $q$, $u$) terms in the free energy.
$\Rkk$ and $\Ukk$ are dynamic self-correlations.
They depend only on the difference $\abs{k-k'}$ thanks to their translational invariance in the Trotter space.
Conversely, $q$ and $u$ are their static counterparts.
Moreover, let us emphasize that $\varDelta \propto~\Lr{\avg{\hat{P}_{i}}}_{J}$
can be associated with the superfluid order parameter [Eq.~\eqref{eq:defD}],
while $q \propto~\Lr{\avg{\hat{P}_{i}}^{2}}_{J}$ with the Edwards-Anderson one [Eq.~\eqref{eq:defq}].
We do not associate here a physical interpretation with the variables $\Ukk$ and $u$.
Nevertheless, they correlate with the onset of superfluidity, as they break the $U(1)$ symmetry.

\subsection{Numerical calculations}
We solve the self-consistent equations \eqref{eq:selfcon_R}--\eqref{eq:selfcon_u} numerically
by iteratively recalculating the values of $\LR{\Rkk}$, $\LR{\Ukk}$, $\varDelta$, $q$ and $u$,
each time using the previous set of those variables as the parameters in the Hamiltonian.
To calculate the averages, we need to trace over all possible configurations of
$\LR{p_1,p_2,\ldots,p_M;q_1,q_2,\ldots,q_M}$.
As the sets of eigenvalues of operators $\hat{P}$ and $\hat{Q}$ are infinite,
we limit the calculations to up to $2$ particles per site,
resulting in $\hat{P}$ and $\hat{Q}$ each having $3$ eigenvalues
in the truncated basis.
This simplification, albeit quite radical,
still leaves as many as $3^{2M}$ possible configurations in the trace.
We cannot use the Monte Carlo method here,
as the discussed model has a severe sign problem \cite{Piekarska2019_APPA135}.
This limits the range of numerically available values of $M$ to just a few.
Here, the majority of the results are obtained with $M=5$.
To compute the integrals over $x_P$, $x_Q$, and $x_B$,
we employ the Gauss quadrature method of integration \cite{Greenwood1948_BotAMS54}.
While the latter is highly efficient and allows determining the integral up to acceptable precision with just a few points,
it increases the computational cost by $1$--$3$ orders of magnitude
compared to this model but with $J_0=0$ and $q=0$ \cite{Piekarska2018_PRL120}.

\section{Critical lines}\label{sec:crit}
We are dealing here with two order parameters, $q$ and $\varDelta$.
Let us recall that the phases of the system that are distinguishable based on these parameters are: disordered ($q = 0$, $\varDelta = 0$),
glass ($q > 0$, $\varDelta = 0$) and superfluid ($q > 0$, $\varDelta > 0$).
With Eqs.~\eqref{eq:selfcon_D} and \eqref{eq:selfcon_q} in hand, one can see why
a phase with $q = 0$ and $\varDelta > 0$ is impossible within the used model:
$\varDelta \neq 0$ requires $\avg{p_{k}} \neq 0$ for some choice of $x_{P}$, $x_{Q}$ and $x_{B}$,
which would in turn imply $q \neq 0$ as well.

In this section, we establish the methods for determining the critical lines.
The boundaries between the abovementioned three regions can be found from the Landau theory of phase transitions.
We formulate them in Sec.~\ref{ssec:crit:land} and analyze in Sec.~\ref{ssec:crit:gen}.
In the following, we also show that regions characterized by long-range superfluid order
divide into the regular SF phase and a superglass phase.
In the latter, on top of the long-range order,
we additionally deal with a glassy one that manifests itself by breaking of the replica symmetry.
Based on it, in Sec.~\ref{ssec:crit:sg} we describe the condition used by us for recognizing the SG phase from the SF one.

\subsection{Landau theory conditions}\label{ssec:crit:land}
The transition between disordered and glass phases occurs in the absence of $\varDelta$,
so it can be found based on the usual Landau theory with a single order parameter,
but the other two transitions need careful handling.
Upon writing down the two-order-parameter free energy
and explicitly evaluating the minima conditions in all phases \cite{Deutges1980_pssb101},
we arrive at the following set of conditions:
\begin{itemize}
	\item if $(\mfrac{\partial^2 F}{\partial \varDelta^2})\big|_{\varDelta=0} < 0$ then the phase is superfluid,
	\item otherwise, the sign of $(\mfrac{\partial^2 F}{\partial q^2})\big|_{q=0}$
	distinguishes between glass (negative) and disordered (positive) phases.
\end{itemize}
Note, that
\begin{itemize}
	\item $(\mfrac{\partial^2 F}{\partial \varDelta^2})\big|_{\varDelta=0}$ is evaluated without setting $q = 0$,
	\item the sign of $(\mfrac{\partial^2 F}{\partial q^2})\big|_{q=0}$ has no useful meaning
	when $(\mfrac{\partial^2 F}{\partial \varDelta^2})\big|_{\varDelta=0} < 0$.
\end{itemize}
In the replica-symmetric model considered here,
$(\mfrac{\partial^2 F}{\partial q^2})\big|_{q=0}$ has the opposite sign than it would normally have \cite{Binder1986_RoMP58},
and thus we need to swap the sides in the glass transition condition.
The final condition is that
$(\mfrac{\partial^2 F}{\partial q^2})\big|_{q=0} < 0$ occurs in the disordered phase,
while $(\mfrac{\partial^2 F}{\partial q^2})\big|_{q=0} > 0$ indicates the glass ordering.

We evaluate $(\mfrac{\partial^2 F}{\partial q^2})\big|_{q=0}$ and $(\mfrac{\partial^2 F}{\partial \varDelta^2})\big|_{\varDelta=0}$
for the free energy from Eq.~\eqref{eq:FreeEnergy},
and arrive at a condition for the glass transition given by
\begin{multline}
	\!\!\!1 = \frac{J^2\beta^2}{2M^4} \sum_{kk'll'} \DxDy \\\times\Lr[\Big]{
		\avg{p_{k}p_{k'}}\avg{p_{l}p_{l'}} - 4\avg{p_{k}p_{k'}}\avg{p_{l}}\avg{p_{l'}} + 3\avg{p_{k}}\avg{p_{k'}}\avg{p_{l}}\avg{p_{l'}}\\
		+2\avg{p_{k}q_{k'}}\avg{p_{l}q_{l'}} - 8\avg{p_{k}q_{k'}}\avg{p_{l}}\avg{q_{l'}} + 6\avg{p_{k}}\avg{q_{k'}}\avg{p_{l}}\avg{q_{l'}}\\
		+\avg{q_{k}q_{k'}}\avg{q_{l}q_{l'}} - 4\avg{q_{k}q_{k'}}\avg{q_{l}}\avg{q_{l'}} + 3\avg{q_{k}}\avg{q_{k'}}\avg{q_{l}}\avg{q_{l'}}
	},
\end{multline}
while the superfluid transition condition reads
\begin{multline}
	\!\!\!1 = \frac{J_0\beta}{2M^2}\sum_{kk'} \DxDy \\\times\Lr[\Big]{
		\avg{(p_{k}+q_{k})(p_{k'}+q_{k'})}
		-\avg{p_{k}+q_{k}}\avg{p_{k'}+q_{k'}}
	}.
\end{multline}
The derivation of the above two equations can be found in Appendices \ref{app:q} and \ref{app:Delta}, respectively.

\subsection{General predictions}\label{ssec:crit:gen}
When approaching either of the disordered-ordered transitions from the disordered phase,
we can evaluate the critical conditions setting both $q = 0$ and $\varDelta = 0$ simultaneously.
In such a case, the conditions simplify, and we arrive at
\begin{equation}\begin{split}
	\!\!\!\mbox{GL:~~} 0 < \frac{\partial^2 F}{\partial q^2}\Bigr|_{\substack{q=0\\\varDelta=0}} = -\lr*{\JBM}^2M^2 + \lr*{\JBM}^4 \lr*{\sum_{kk'}\Rkk}^2,\\
	\!\!\!\mbox{SF:~~} 0 < \frac{\partial^2 F}{\partial \varDelta^2}\Bigr|_{\substack{q=0\\\varDelta=0}} = 2\lr*{\JOBM}M - 2\lr*{\JOBM}^2 \lr*{\sum_{kk'}\Rkk},
\end{split}\end{equation}
which can be written in the form
\begin{equation}\begin{split}
	\!\!\!\mbox{GL:~~~} J^2\mathcal{A}^2 & > 1,\\
	\!\!\!\mbox{SF:~~~~~} J_0\mathcal{A}   & < 1,
\end{split}\end{equation}
with the same $\mathcal{A} = \mfrac{\beta \sum_{kk'}\Rkk}{M^2}$
in both cases.
The essential condition for the DI-GL transition is $J^2\mathcal{A}^2 = 1$,
however, we also need $J_0\mathcal{A} < 1$ as we should not be in the SF phase.
These two conditions combined indicate that the DI-GL transition can occur only when $J>J_0$.
Analogously, one can find that the DI-SF transition can only occur when $J<J_0$.
Thus, when leaving the DI phase,
we can uniquely determine which phase we enter
based only on the relation of $J$ to $J_0$.
In particular, we can also conclude that in the vicinity of the disordered phase region,
the SF-GL transition takes place at $J=J_0$.

Moreover, when $\varDelta = 0$,
the Hamiltonian is independent of $J_0$ [see Eq.~\eqref{eq:effham}].
As a consequence, the DI-GL transition is independent of $J_0$ as well,
as neither of the phases on the two sides has $\varDelta > 0$.

\subsection{Superglass phase}\label{ssec:crit:sg}
We expect that the phase we described above as superfluid consists, in fact, of two parts:
one is the usual superfluid phase with long-range order,
while the other one is a superglass (SG) phase and
exhibits both superfluid and glassy orderings simultaneously.
However, the implication $(\varDelta > 0) \Rightarrow (q > 0)$ means that in the superfluid phase,
$q$ is no longer a measure of glass order.
Thus, using the order parameters defined so far,
we cannot distinguish between these two.

Two ways of recognizing the glass order on top of the superfluid one were used by other authors.
The first one is to find the point of breaking the replica symmetry~\cite{Yu2012_PRB85},
while the second one requires checking translational symmetry breaking~\cite{Carleo2009_PRL103}.
We focus on the first of these methods.

The derivation of the stability condition for the replica-symmetric solution,
similar to the one done by de Almeida and Thouless~\cite{Almeida1978_JoPAMaG11} but more complex,
may be found in Ref.~\cite{Piekarska2022a}.
There, compared to the spin glass case, we deal with more independent variables
(five deviations from the symmetric solution instead of two) and additional Trotter dimensions,
which results in matrix elements becoming matrix blocks.
We follow the standard treatment of checking positive-semidefiniteness of the Hessian matrix of the free energy,
in which we additionally Fourier transform to get rid of one of the Trotter dimensions~\cite{Buettner1990_PRB41},
and arrive at a set of conditions for the stability.
Since it is not straightforward to determine any of them as redundant,
we numerically investigate which of them is the strongest.
This treatment allows us to arrive at a condition that the solution is unstable if and only if
there is a negative eigenvalue of a certain matrix,
which is a matrix analog of the ``$P-2Q+R$'' eigenvalue in spin glasses~\cite{Almeida1978_JoPAMaG11}.

Here, we employ the described condition derived in Ref.~\cite{Piekarska2022a} as a way to detect the glassy order.
It allows us to numerically find the transition line based on averages
of various combinations of $\hat{P}$ and $\hat{Q}$ (up to $4$ operators).
We note that such treatment does not indicate the superglass phase explicitly,
as the described stability transition exists both in the $\varDelta=0$ and $\varDelta\neq 0$ areas.
In the non-superfluid part of the phase diagram, it coincides with the DI-GL transition condition from the Landau theory.
However, in the $\varDelta\neq 0$ region, it gives rise to a new critical line,
which is the sought transition to the SG phase.
We use the latter as the SF-SG critical line.

\section{Results}\label{sec:results}
In this Section, we present numerical results
obtained from solving the self-consistent equations \eqref{eq:selfcon_R}-\eqref{eq:selfcon_u}
and evaluating the critical-line conditions from Sec.~\ref{ssec:crit:land} and Sec.~\ref{ssec:crit:sg}.
We begin in Sec.~\ref{ssec:results:gen}
with qualitative considerations of the phase transitions found in the system.
Then, in Sec.~\ref{ssec:results:detail},
we quantitatively describe constant-temperature phase diagrams
and analyze the phases.
Finally, in Sec.~\ref{ssec:results:temp},
we discuss the impact of temperature
and extrapolate our results to $M\to\infty$,
which allows us to predict quantum phase transitions in the system.

\subsection{General discussion}\label{ssec:results:gen}
\begin{figure}[tb]
	\includegraphics[width=\columnwidth]{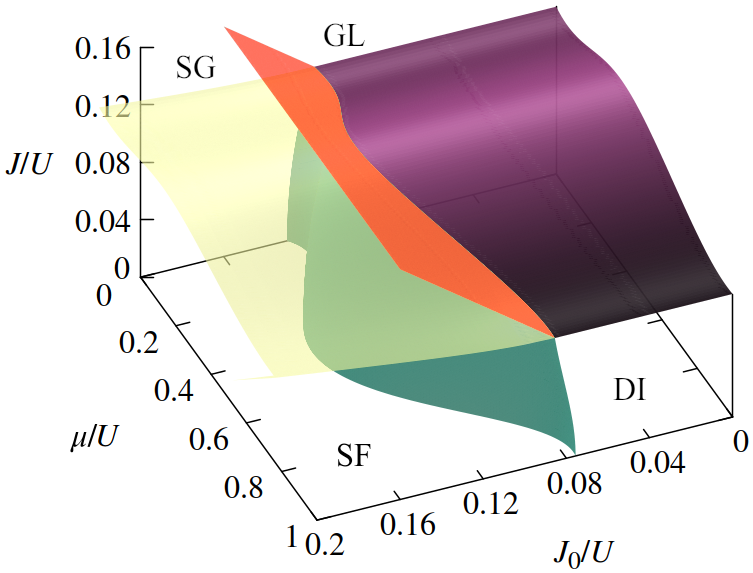}
	\caption{Phase diagram in variables $\mu/U$--$J_0/U$--$J/U$ at $T/U = 0.08$.
	Color of the surfaces indicates the transition that takes place.
	The purple one separates disordered (bottom) from glass (top);
	red: superglass (left) from glass (right);
	yellow: superfluid (bottom) from superglass (top);
	cyan: superfluid (left) from disordered (right).}
	\label{fig:diag3D}
\end{figure}

We begin with presenting in Fig.~\ref{fig:diag3D}
a 3-dimensional phase diagram of the studied system in the $\mu/U$--$J_0/U$--$J/U$ space of parameters.
It is obtained at $T/U = 0.08$ and $M = 5$.
On the diagram, we can find all four phases, separated by the plotted surfaces.
The purple surface illustrates the Landau-theory glass transition condition from Sec.~\ref{ssec:crit:land}
and separates the disordered phase ($q = 0$, $\varDelta = 0$; below the surface) from the glass phase ($q > 0$, $\varDelta = 0$; above the surface).
The red and cyan surfaces are Landau-theory superfluid transition conditions,
with the red separating superglass ($q > 0$, $\varDelta > 0$, broken replica symmetry; on the left) from glass (on the right)
while the cyan one separates superfluid ($q > 0$, $\varDelta > 0$; left) from disordered (right).
The yellow surface stems from the stability condition from Sec.~\ref{ssec:crit:sg}
and separates superfluid (bottom) from superglass (top).

As predicted in Sec.~\ref{ssec:crit:gen},
the surface between DI and GL does not depend on $J_0$.
The SF-SG transition approximately retains this shape as well.
The SG/SF-GL transition next to the disordered lobe occurs at $J=J_0$,
again following our prediction.
Deeper inside the ordered phase, the same argument cannot be used, so we have no strict proof, 
but the numerically found SF-GL transition remains close to the $J=J_0$ plane.
However, within the current model and computational resources, we cannot determine
whether $J=J_0$ is the exact transition point.
In spin glasses, $J=J_0$ was the point of the ferromagnetic-glass transition~\cite{Binder1986_RoMP58}.
Therefore, we do not find it unlikely that the approximate dependence we observe is, in fact, exact.

We find that all four phases meet at a four-critical line.
There are no direct SF-GL and SG-DI transitions
except at the four-critical line.

Having established the qualitative picture of the discussed phases,
we turn to a more detailed analysis of the phase diagram by inspecting its various cross-sections.

\subsection{Detailed phase diagrams}\label{ssec:results:detail}
\begin{figure}[tb]
	\includegraphics[width=\columnwidth]{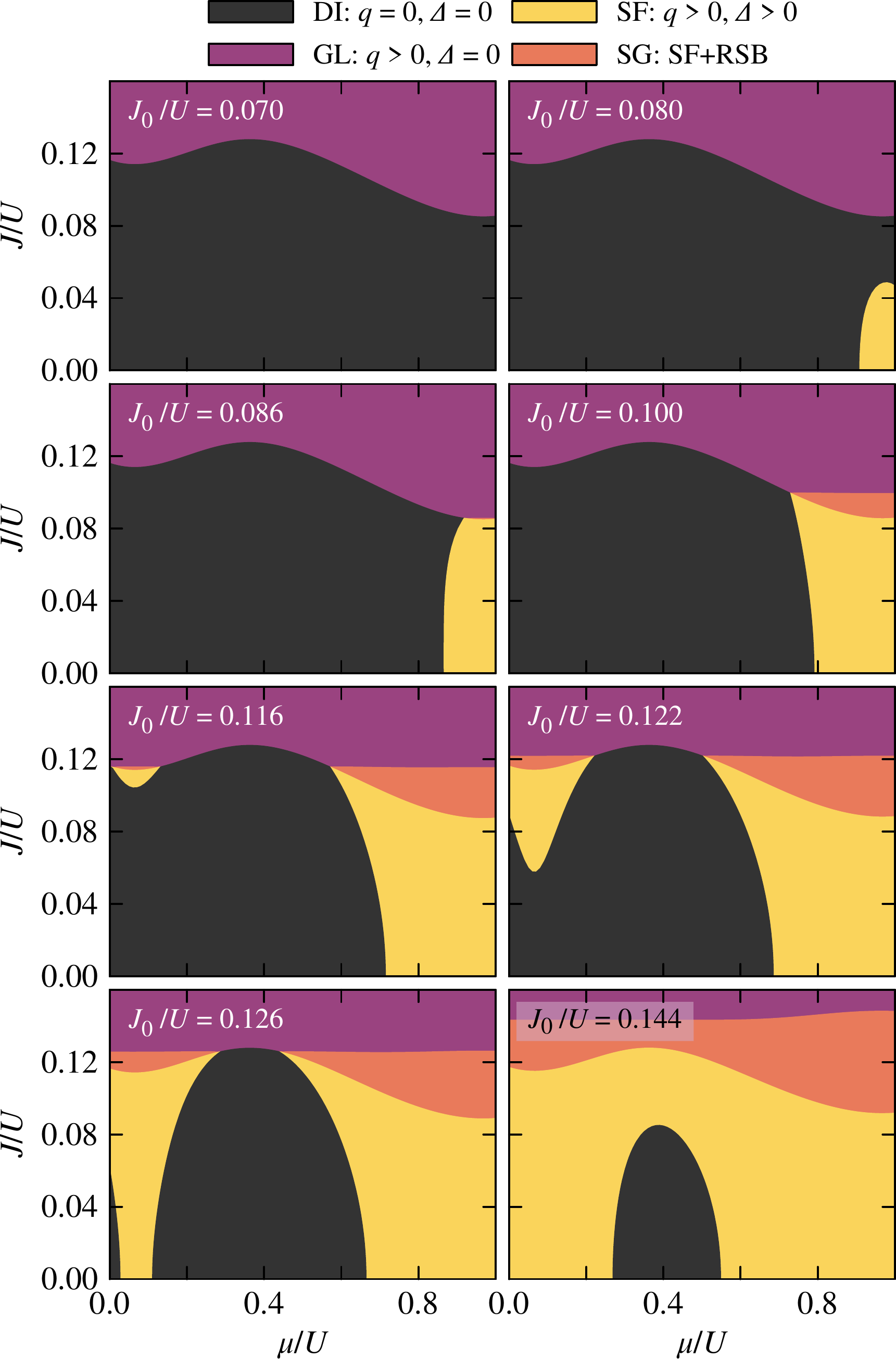}
	\caption{Phase diagrams in variables $\mu/U$--$J/U$ at $T/U = 0.08$
	and at various values of $J_0/U$ (given on each of the panels).}
	\label{fig:muJ}
\end{figure}

First, in Fig.~\ref{fig:muJ}, we plot a sequence of diagrams
obtained at increasing fixed values of $J_0/U$.
The first ($J_0/U = 0.07$) is analogous to the one for $J_0/U = 0$
found in Ref.~\cite{Piekarska2018_PRL120}
since it contains only DI and GL phases,
and the transition between them does not depend on $J_0$.
As $J_0$ is increased, around $J_0/U \approx 0.076$, the SF phase appears
in the $J<J_0$ region at the expense of the DI phase.
Its areas on both sides of the DI lobe grow and join so that 
around $J_0/U \approx 0.128$,
the DI phase no longer has a direct transition to the GL phase.
At the same time, a layer of the SG phase separating GL and SF phases emerges as $J_0$ is increased.
Finally, around $J_0/U \approx 0.152$,
the DI phase vanishes completely.
Note that the analyzed results are calculated at finite $M$,
so the exact values of $J_0/U$ thresholds will be different
in the $M\to\infty$ limit,
but the qualitative features will stay the same.

As mentioned earlier,
the SF-SG transition line closely resembles the DI-GL line found at lower values of $J_0/U$.
There is a difference between these two,
noticeable close to integer $\mu/U$ values,
which indicates a weak $J_0$-dependence.
It could also result from the inaccuracy of the calculation method,
possibly more pronounced within ordered phases.
However, both lines evolve similarly with $M$.
They are also virtually independent of integration stencil,
which we consider the most likely source of errors in ordered phases.
Nonetheless, this similarity is understandable
as a result of introducing superfluid order by nonzero mean interaction
to areas with the presence and absence of glassy order.
In the SG phase, these two orders compete~\cite{Yu2012_PRB85}.
We study this in detail below.

\begin{figure}[tb]
	\includegraphics[width=\columnwidth]{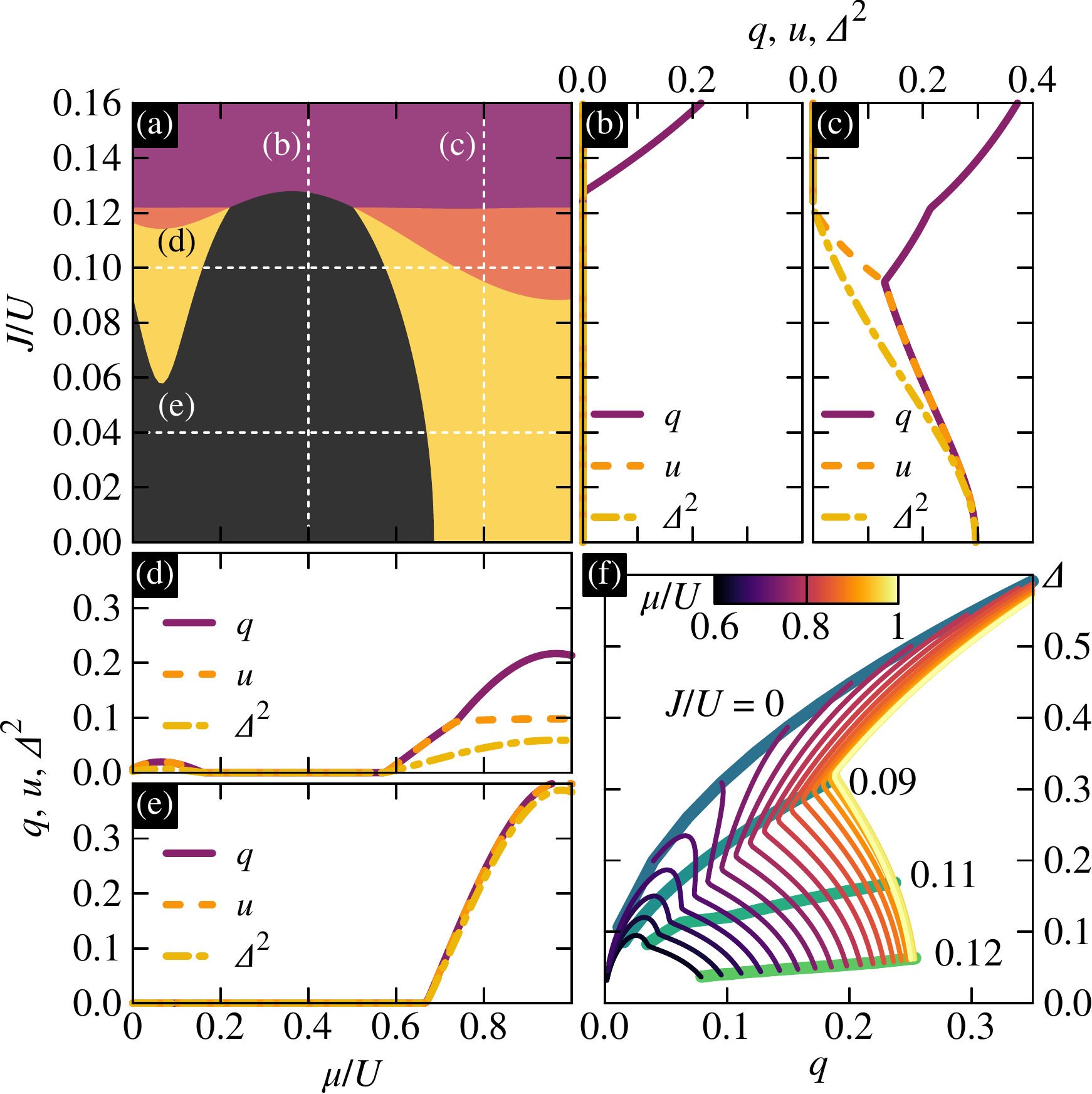}
	\caption{(a) Phase diagram at $J_0/U=0.122$ repeated from Fig.~\ref{fig:muJ}.
	(b-e) Values of order parameters
	along cuts of the phase diagram from panel (a), as marked.
	The cuts are at (b) $\mu/U=0.4$ (c) $\mu/U=0.8$ (d) $J/U=0.10$ (e) $J/U=0.04$.
    Note that on panels (b) and (c) the y-axis is common with panel (a),
	while the value of the order parameters is at the x-axis.
	(f) Correlation between $q$ and $\varDelta$ in the $\mu/U>0.6$ part of superfluid.
	Points with the same $\mu/U$ are connected with thin lines
	and color-coded with black-purple-yellow gradient, with values given by the color bar,
	while points with the same $J/U$ are connected with thick lines
	and color-coded with green-blue gradient, as labeled on the plot.}
	\label{fig:cuts}
\end{figure}

Let us focus on the order parameters.
In Fig.~\ref{fig:cuts}, we analyze values of $q$, $\varDelta$, and $u$
for the phase diagram at $J_0/U = 0.122$.
Panel (a) is repeated from the previous Figure
and serves as a key for the other panels.
In panels (b) and (c), we plot values of $q$, $u$, and $\varDelta^2$
along vertical cuts at $\mu/U=0.4$ and $\mu/U=0.8$, as marked in panel (a).
Note that the $y$ axis is common with panel (a),
while the value of the order parameters is at the $x$-axis, i.e., the plots are rotated clockwise.
In panel (b), we observe a transition between disordered and glassy phases.
As expected, both these phases have $u=0$ and $\varDelta=0$,
while $q$ continuously goes from $0$ on the DI side
to a finite value on the GL side.
Panel (c) shows a transition from superfluid through superglass to glass phase.
The order parameters $q$ and $\varDelta$ behave as expected.
Namely, $q$ stays nonzero across all phases,
while $\varDelta$ (and subsequently $u$) has a finite value in the SF and SG phases
and vanishes when the phase changes to a glass.

However, we did not have a prediction for the quantitative relative behavior of these variables.
Here, we find that in the superfluid phase, $u=q$,
and they both decrease as $J/U$ increases.
Then, when entering the superglass phase, they split:
$q$ starts increasing and is no longer equal to $u$,
which keeps further decreasing.
Both values have a cusp at this point.
Finally, when $J$ reaches the glass transition point, $u$ vanishes,
while $q$ has another cusp but keeps increasing.
One may interpret this result by associating the nonzero value of $q-u$
with the existence of the glass order on top of the superfluid order.
Such subtraction of a superfluid ``base'' from the regular Edwards-Anderson order parameter
remotely resembles the translational-symmetry-breaking superglass condition from Ref.~\cite{Carleo2009_PRL103},
mentioned previously as the other method of determining the superglass transition.

Going back to the analysis of order parameters,
in panels (d) and (e) of Fig.~\ref{fig:cuts},
we plot horizontal cuts of the phase diagram at $J/U=0.04$ and $J/U=0.10$, as marked in the panel (a).
Both these panels show a transition between disordered and superfluid phases.
Panel (d) then goes through the superglass phase,
on the boundary of which one can see a cusp in $q$ and $u$ in this direction as well.
On the panel (f), we plot the correlation between $q$ and $\varDelta$
in the $\mu/U>0.6$ part of the superfluid and superglass phases.
Each calculated point
from the $(\mu/U = 0.6$--$1.0) \times (J/U = 0$--$0.122)$ rectangle
is plotted at corresponding coordinates $(q, \varDelta)$.
Points with the same value of $\mu/U$ are connected with thin lines
and color-coded with a black-purple-yellow gradient.
Points with equal values of $J/U$ are connected with thick lines color-coded with a green-blue gradient and labeled.
Note that the $J/U$ steps between these last lines are not equal.
We notice that at a constant $J/U$ but varying $\mu/U$,
values of $q$ and $\varDelta$ are always correlated with each other (positive slope).
However, when $\mu/U$ is constant but $J/U$ varies, one can see
both correlated behavior at lower $J/U$, and anticorrelated one (negative slope) at higher $J/U$.
Notably, the two coincide with SF and SG areas, respectively.
This illustrates the competition of the two types of ordering coexisting in the SG phase.

\begin{figure}[tb]
	\includegraphics[width=\columnwidth]{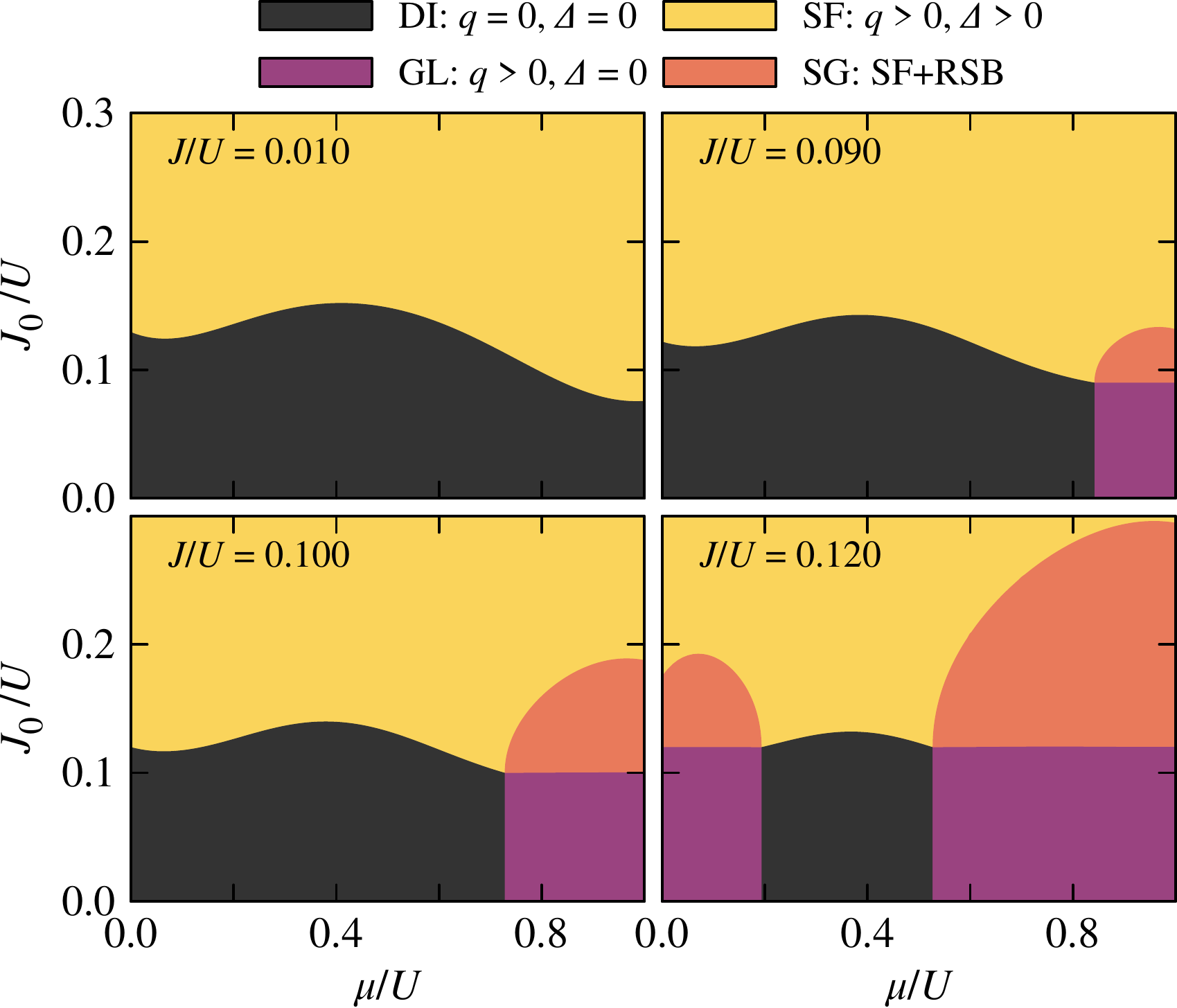}
	\caption{Phase diagrams in variables $\mu/U$--$J_0/U$ at $T/U = 0.08$
	and at various values of $J/U$ (given on each of the panels).}
	\label{fig:muJ0}
\end{figure}

Cutting the phase diagram along constant $J/U$,
we make a link to the widely studied non-disordered bosonic systems \cite{Lewenstein2007_AiP56}.
In Fig.~\ref{fig:muJ0}, we present phase diagrams in the same variables
as in the usual Bose-Hubbard model, namely $\mu/U$ and $J_0/U$
(which in the works concerning the non-disordered case is denoted as just $J/U$).
The first panel, obtained for a nearly zero disorder,
can be, in fact, regarded as the no disorder limit.
In Appendix~\ref{app:compare}, we show a corresponding direct comparison
with Ref.~\cite{Stasyuk2009_CMP12} that shows quantitative agreement.
The subsequent panels present the phase diagrams at increasing levels of disorder.
With the latter, at around $J/U\approx 0.086$, the glass and superglass phases appear at high $\mu/U$.
Further increasing $J$ makes the glassy phases extend,
taking the place of the disordered and superfluid phases, respectively.
Finally, around $J/U\approx 0.128$, the disordered phase vanishes.
Unlike the diagonal case~\cite{Fisher1989_PRB40,Bissbort2010_PRA81,Weichman2008_PRB77,Krueger2009_PRB80}
where the Bose Glass phase emerged at the interface between DI and SF phases,
here, the GL phase does not affect the DI-SF line,
as it appears next to these two phases.

\subsection{Temperature dependence}\label{ssec:results:temp}
\begin{figure}[tb]
	\includegraphics[width=\columnwidth]{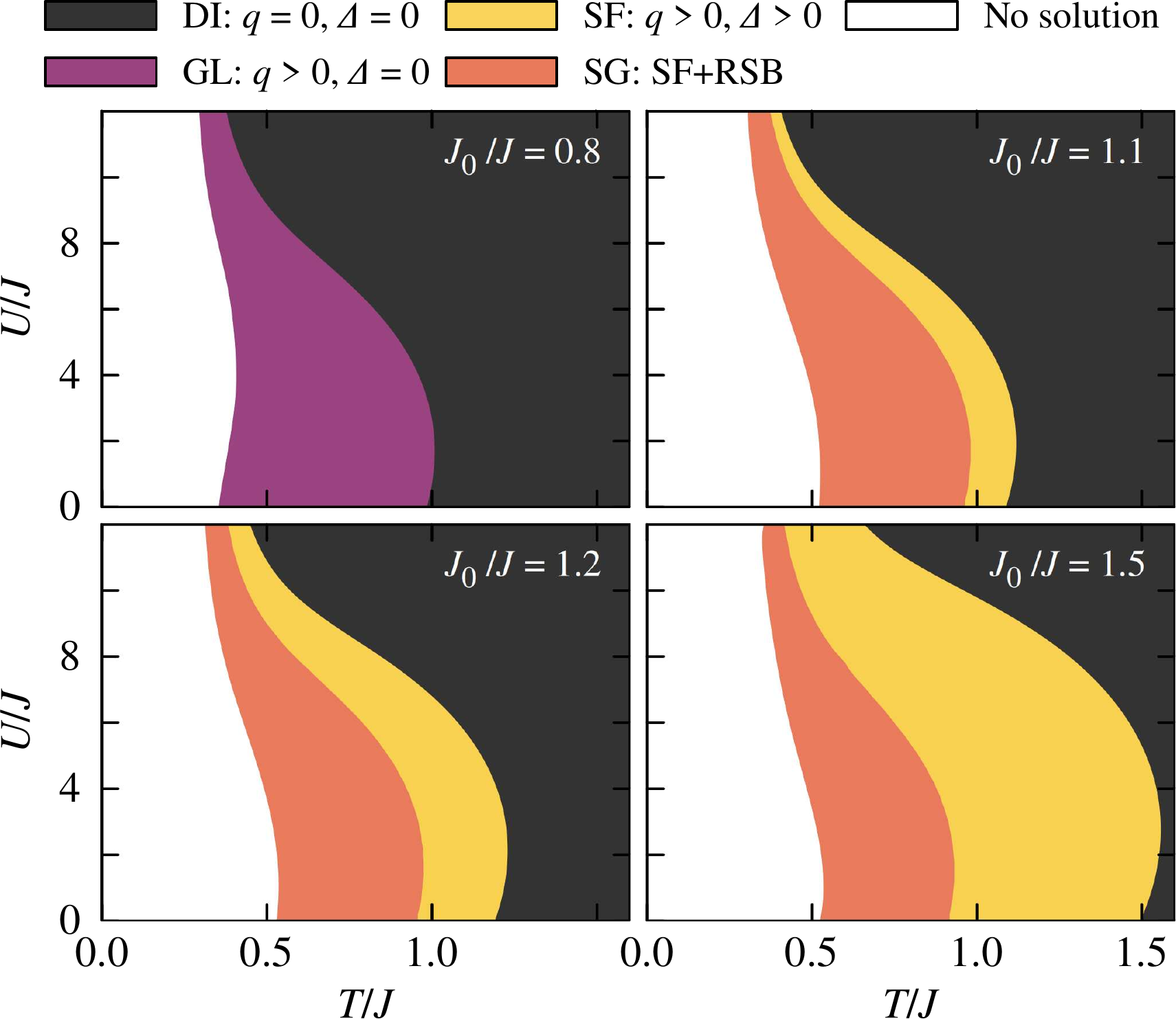}
	\caption{Phase diagrams in variables $T/J$--$U/J$ at $\mu/U = 0.4$
	and at various values of $J_0/J$ (given on each of the panels).}
	\label{fig:TU}
\end{figure}

Finally, we analyze the impact of temperature on the phase diagram.
For this, we choose a fixed value of $\mu/U=0.4$.
To relate to spin-glass works, we use $J$ as an energy scale in this section.

In the first panel of Fig.~\ref{fig:TU},
we show the phase diagram in variables $T/J$--$U/J$ at $J_0/J = 0.8$.
Only the disordered and glass phases are present
within the computationally accessible area
(inaccessible points are left blank).
Under the assumption that the SF-GL boundary is at $J_0\approx J$,
there will be no superfluid phase at $J_0\lesssim J$ and, therefore,
all the $J_0/J\lesssim 1$ phase diagrams will be the same as the depicted $J_0/J = 0.8$ one
since the GL-DI transition does not depend on $J_0$.
In particular, this phase diagram is also analogous to the one at $J_0=0$,
which we discussed previously~\cite{Piekarska2018_PRL120}.
On the other hand,
there is no glass phase in the $J_0>J$ regime shown in the remaining three panels,
as the long-range order has appeared at the anticipated $J_0\approx J$ phase transition.
The disordered phase boundary is not constant anymore either.
One can see the superfluid phase growing as $J_0/J$ increases.

\begin{figure}[tb]
	\includegraphics[width=\columnwidth]{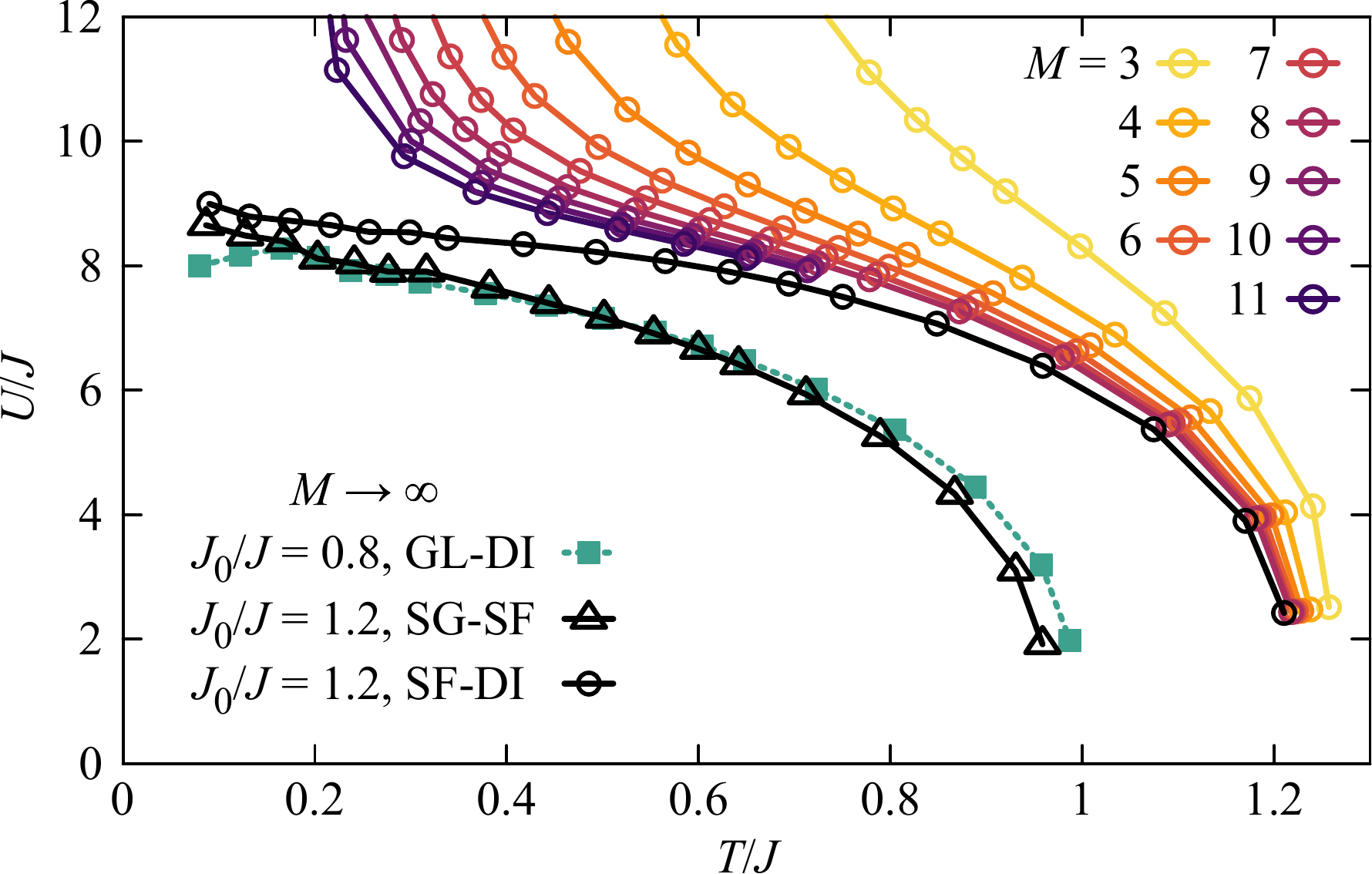}
	\caption{Critical lines in variables $T/J$--$U/J$ at $\mu/U = 0.4$ and $J_0/J = 1.2$:
		SF-DI for $M=3$--$11$ (yellow to purple, empty circles);
		their extrapolation to $M=\infty$ (black, empty circles);
		SG-SF extrapolation (black, empty triangles);
		GL-DI extrapolation at $J_0/J = 0.8$ (cyan, full squares).
		Lines are to guide the eye only.}
	\label{fig:extrapol}
\end{figure}

In the previous sections, we dealt with results obtained at relatively high temperatures, 
where the finite value of $M$ does not significantly influence the results.
However, as we decrease the temperature,
the method used by us is expected to be too inaccurate once $T/J\lesssim 1/M$.
In order to decrease the lower bound of feasible temperatures,
we extrapolate the finite-$M$ critical lines to $M\to\infty$.
Following the analysis from Ref.~\cite{Suzuki1985_PLA113},
a thermal average of an observable $X$ should scale as
\begin{equation}
	\avg{X_M} = \frac{a}{b+M^2} + \avg{X_\infty},
\end{equation}
where $\avg{X_\infty}$ is the converged result, while the term with parameters $a$ and $b$ is the error resulting from finite $M$.
In Fig.~\ref{fig:extrapol}, we plot a set of DI-SF critical lines (empty colored circles)
analogous to the $J_0/J=1.2$ panel of Fig.~\ref{fig:TU} calculated for $M=3$--$11$,
along with the $M\to\infty$ extrapolation of these (empty black circles).
We also plot the $M\to\infty$ extrapolation of the SF-SG line (empty black triangles).
For the comparison, we include $J_0/J=0.8$ extrapolation of the DI-GL critical line (full cyan squares),
which is analogous to the $J_0=0$ extrapolation reported before \cite{Piekarska2018_PRL120}.
We find that all extrapolated lines approach a finite value of $U/J$ as $T\to 0$.
Thus, we deal with quantum phase transitions in all cases.

\begin{figure}[tb]
	\includegraphics[width=\columnwidth]{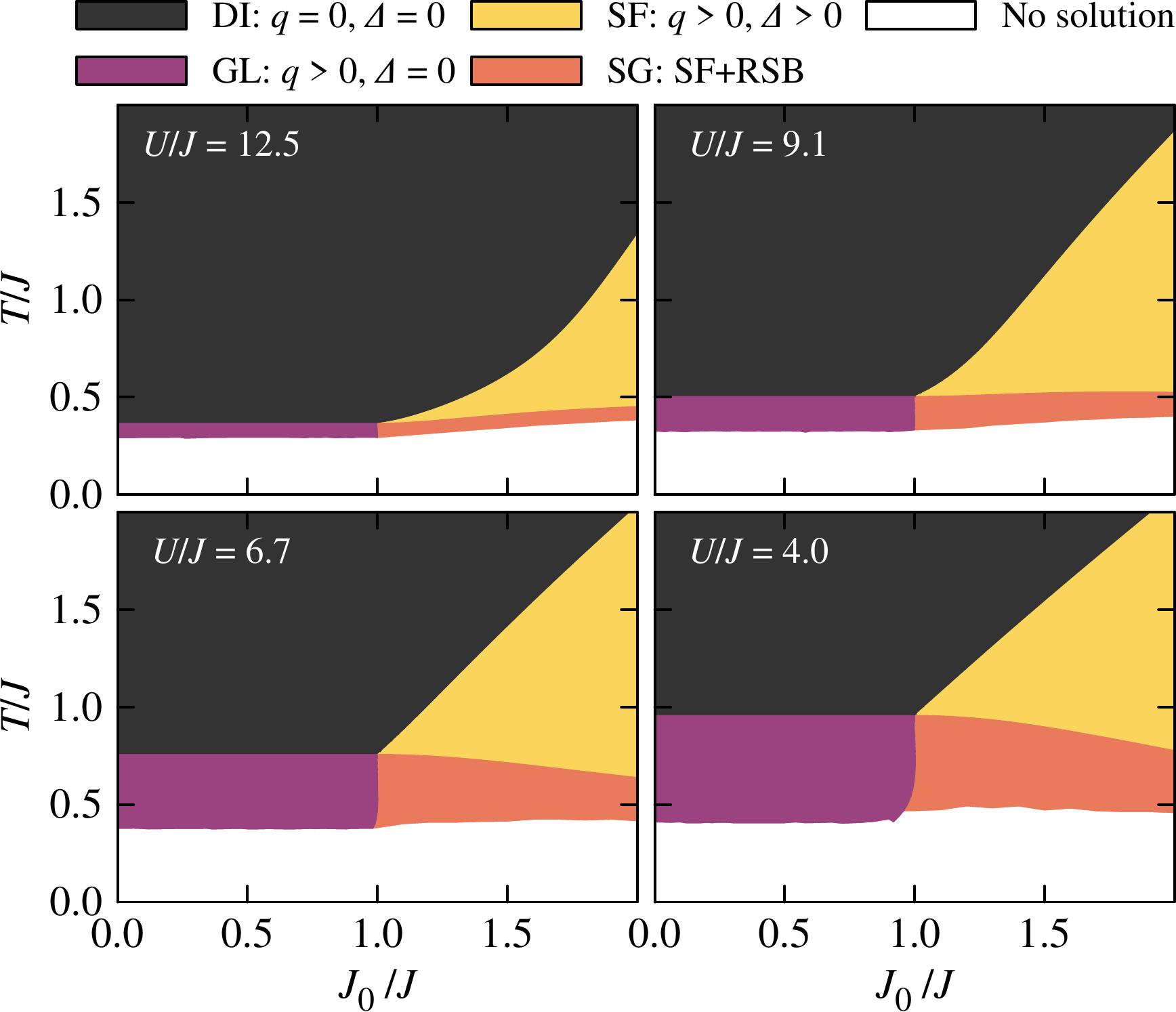}
	\caption{Phase diagrams in variables $J_0/J$--$T/J$
	at $\mu/U = 0.4$ and at various values of $J/U$ (given on each of the panels).}
	\label{fig:J0T}
\end{figure}

The four panels of Fig.~\ref{fig:J0T}
show the phase diagram at various values of $U/J$.
All four qualitatively resemble analogous diagrams
obtained in the spin-glass systems~\cite{Sherrington1975_PRL35,Pirc1985_ZfPBCM61}.
The main differences are threefold:
First, in place of the ferromagnetic phase found in spin systems, we deal with two phases with superfluid order: SF and SG. 
Second, in the upper panels, the DI-SF line is not linear.
Third, at low temperatures, the SG-GL line bends towards $J_0<J$.

The DI-SF line nonlinearity can be attributed to a finite value of $M$.
Checking the extrapolation in Fig.~\ref{fig:extrapol}, one can see
that, e.g., at the value of $U/J=12.5$, there is a disordered phase at $T\to 0$.
However, it does not manifest itself in any finite $M$.
This means that the existence of the glass phase and DI-GL transition
in the top panels is an effect of finite $M$.
Thus, the DI-SF line near $J_0=J$ is far from the real one,
as its starting point is at $T=0$.
The bending of the GL-SG line is also a result of numerical insufficiency.
The extrapolation to $M\to\infty$ combined with employing a more demanding integration stencil
suggest that the line in fact bends slightly in the other direction ($J_{0}>J$).
This is in line with the replica-symmetric spin-glass result~\cite{Sherrington1975_PRL35}.
There, breaking of the replica symmetry resulted in the line at exactly $J_{0}=J$~\cite{Binder1986_RoMP58},
which we envisage to be the result also here.

The last panel not only recovers the spin-glass result qualitatively,
but it is quantitatively quite close as well.
In the $U/J\to 0$ limit, the lines present in the spin-glass phase diagram would be recovered exactly.

In the discussion of Fig.~\ref{fig:muJ}, we have noticed a similarity between DI-GL and SF-SG lines.
Here, based on diagrams in Fig.~\ref{fig:J0T}, we may study it in more detail.
If the two curves coincided, the SF-SG line
would be just a straight horizontal continuation of the DI-GL line.
However, the line bends,
meaning that the difference between the two curves increases with increasing $J_{0}$.

\section{Summary}\label{sec:summary}
We have studied a many-body system of disordered interacting bosons
by modeling it using the Bose-Hubbard Hamiltonian with a random hopping term.
The averaging has been done exactly,
i.e., not as an average of some realizations of the disorder,
but as an analytical integration over the entire disorder distribution.
The major advance over previous works
is taking into account a possibility of a nonzero mean of the hopping distribution.
We have analytically derived critical line conditions in such a setting,
followed by numerically obtaining phase diagrams spanned across various sets of parameters.

We have distinguished four phases:
a high-temperature disordered phase,
a superfluid phase characterized by a long-range order,
a glassy phase characterized by the Edwards-Anderson order parameter,
and a superglass phase where both these orders coexist.
Upon analyzing the behavior of the order parameters in the superglass phase,
we have found that the two orders compete (anticorrelate) within this phase.
We have shown and analyzed phase diagrams of the system
and, where it was possible, compared them to those of spin-glasses, diagonally disordered and non-disordered systems.

As proposed in Ref.~\cite{Piekarska2018_PRL120}, an experimental implementation of the studied model is possible
with a fully connected disordered wood-pile arrangement of elongated optical traps linked via Josephson junctions.
Thus, there are prospects for verifying our results and, in particular, observing the emergent superglass phase.

\begin{acknowledgments}
This work was supported by the Polish National Science Centre under Grant No 2018/31/N/ST3/03600.
Calculations have been partially carried out using resources provided by Wroclaw Centre for Networking and Supercomputing (\url{http://wcss.pl}), grant No. 449.
\end{acknowledgments}

\appendix
\section{Zero disorder limit}\label{app:compare}
\begin{figure}[tb]
	\vspace{0.3cm}
	\includegraphics[width=\columnwidth]{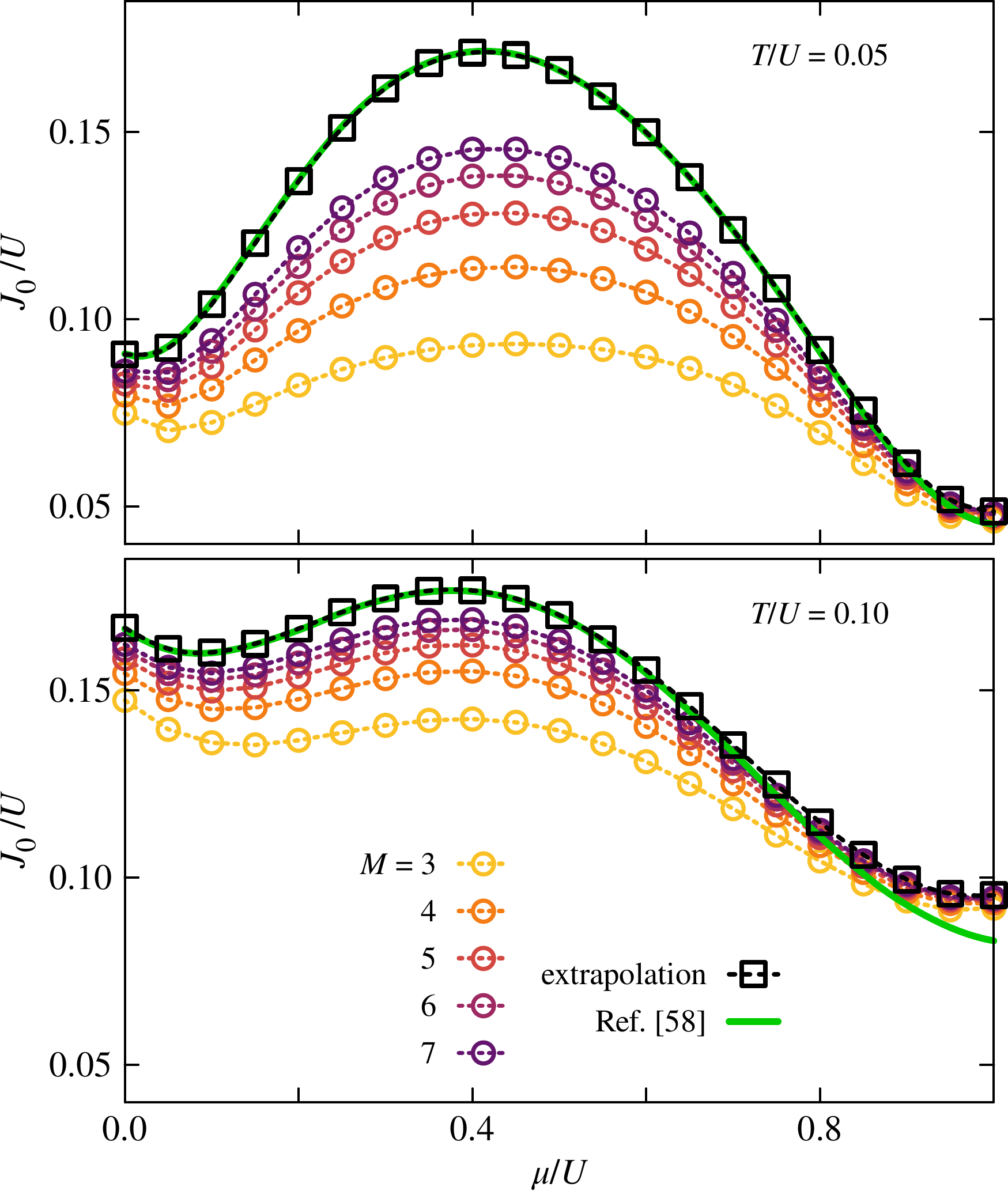}
	\caption{Phase diagram in variables $\mu/U$--$J_0/U$
	at $T/U = 0.05$ (top) and $T/U = 0.10$ (bottom).
	Critical lines in $M=3$--$7$, calculated at $J/U=10^{-4}$,
	are marked with empty yellow-purple circles,
	while their $M\to\infty$ extrapolation is marked with empty black squares.
	Dotted lines are to guide the eye only.
	Green solid line is the zero-disorder critical line from Ref.~\cite{Stasyuk2009_CMP12}.}
	\label{fig:app:cmp}
\end{figure}

In Fig.~\ref{fig:app:cmp} we present a phase diagram obtained in the same manner as Fig.~\ref{fig:muJ0} 
for nearly no disorder ($J/U=10^{-4}$) and at two different temperatures.
The Figure shows finite $M$ data along with their extrapolation to $M\to\infty$.
We compare this result to an earlier work \cite{Stasyuk2009_CMP12},
where a non-disordered Bose-Hubbard model at finite temperatures was studied
in the Hubbard operator formalism using the random phase approximation.
There, the authors found the critical line by looking for the superfluid order parameter becoming nonzero,
which coincided with a divergence in a single-particle Green's function.
We find quantitative agreement at both considered temperatures
in almost entire $\mu/U$ range.
Our solution becomes less reliable at higher temperature and high $\mu/U$.
We attribute this to our cutoff of the basis to two particles per site,
and expect full agreement at higher cutoffs.

\section{Thermodynamic and $n\to 0$ limits}\label{app:limits}
\subsection{Transformation to a single site problem}
We start from the expression for $Z^n$ given in Eq. \eqref{eq:averaged}.
To transform it into a single-site problem,
we apply the Hubbard-Stratonovich transformation to terms containing different site indices.
For example, one kind of terms undergoes
\begin{widetext}
\begin{equation}
	\exp\Lr*{
		\frac{1}{4N}\lr*{
			\JBM\Si \pkia \pkkiaa
		}^2
	}
	= \sqrt{\frac{N}{\pi}} \intid\lkaka^P \exp\Lr*{
		-N(\lkaka^P)^2+\lkaka^P\lr*{
			\JBM\Si \pkia \pkkiaa
		}
	}.
\end{equation}
The full transformed expression reads
\begin{equation}\begin{split}
	\Lr*{Z^n}_{J} ={} &\Trpq\,\Mpq
	\Pka\exp\Lr*{
		-\frac{J_0\beta}{2MN}\sum_{i}\lr*{(\pkia)^2+(\qkia)^2}
	}
	\Pkaka\!\!\exp\Lr*{
		-\frac{J^2\beta^2}{4M^2N}\Si\lr*{(\pkia)^2+(\qkia)^2}\lr*{(\pkkiaa)^2+(\qkkiaa)^2}
	}\\
	&\times \lr*{\sqrt{\frac{N}{\pi}}}^{2n^2M^2}\lr*{\sqrt{\frac{N}{2\pi}}}^{n^2M^2}
	\lr*{\sqrt{\frac{N}{2\pi}}}^{2nM}
	\Pkaka\Lr[\Bigg]{
	    \intid\lkaka^{PQ}
		\exp\lr*{-\frac{N}{2}(\lkaka^{PQ})^2 + \frac{\lkaka^{PQ}J\beta}{M}\Si \pkia \qkkiaa}\\
		&\intid\lkaka^P
		\exp\lr*{-N(\lkaka^P)^2 + \lkaka^P\JBM\Si \pkia \pkkiaa}
		\intid\lkaka^Q
		\exp\lr*{-N(\lkaka^Q)^2 + \lkaka^Q\JBM\Si \qkia \qkkiaa}
	}\\
	&\times\Pka\Lr[\Bigg]{
		\intid\nka^P
		\exp\lr*{-\frac{N}{2}(\nka^P)^2 + \nka^P\sqrt{\JOBM}\Si \pkia}
		\intid\nka^Q
		\exp\lr*{-\frac{N}{2}(\nka^Q)^2 + \nka^Q\sqrt{\JOBM}\Si \qkia}
	},
\end{split}\end{equation}
\end{widetext}
which can be concisely rewritten as
\begin{equation}
	\mathcal{N}\prod_\xi\lr[\Big]{\intid\xi\e^{-N\xi^2}}
	\prod_i\Lr*{
		\Trpq\Mpq^{(i)}\exp\lr*{-\beta\Heff^{(i)}}
	},
\end{equation}
where all constants were incorporated into $\mathcal{N}$,
and $\xi$ runs over all newly introduced variables, i.e.,
\begin{equation}
	\xi \in \LR*{\LR{\lkaka^P}, \LR{\lkaka^Q}, \LR{\lkaka^{PQ}}, \LR{\nka^P}, \LR{\nka^Q}}
\end{equation}
and
\begin{equation}\begin{split}
	-\beta \Heff = &- \frac{1}{N}\LR*{
		\Ska \JBMa{2}\Lr*{(\pka)^2+(\qka)^2}
	}^2\\
	&-\frac{1}{N}\Ska\LR*{
		\JOBM\Lr*{(\pka)^2+(\qka)^2}
	}\\
	&+\sqrt{\JOBM}\Ska\lr*{
		\nka^P\pka+\nka^Q\qka
	}\\
	&+ \JBM\Skaka \lr[\Big]{
		\lkaka^P\pka\pkkaa+\lkaka^Q\qka\qkkaa\\
		&+ \lkaka^{PQ}\pka\qkkaa
	}
	+ \Ska\ln \mathcal{M}_{pq}
\end{split}\end{equation}
is the effective single-site Hamiltonian.
We can now perform the trace over the site indices, obtaining
\begin{equation}
	\Lr*{Z^n}_{J} = \mathcal{N}\prod_\xi\lr[\Big]{\intid\xi}\exp\lr*{-N\mathcal{F}}
\end{equation}
with the effective free energy
\begin{equation}
	\mathcal{F} = \sum_\xi\xi^2 - \ln \Trpq \eBH.
\end{equation}
Note, that $\Trpq(\cdot)$ changed the exact mathematical form here,
but it kept the meaning of being a sum over all possible configurations.

\subsection{Saddle point solution}
In the thermodynamic limit, we use the saddle point method
\begin{equation}
	\intid\lambda\,\mathrm{e}^{-N\mathcal{F}(\lambda)}
	\approx \mathrm{e}^{-N\mathcal{F}(\lambda_0)},
	\quad\mbox{where }\,\frac{\partial\mathcal{F}}{\partial \lambda}\Biggr|_{\lambda=\lambda_0}\!\!\!\!\!\!=0,
\end{equation}
which gives us the set of self-consistent equations
\begin{subequations}
\begin{eqnarray}
	\lkaka^P &=& \JBMa{2}\avg{\pka \pkkaa},\\
	\lkaka^Q &=& \JBMa{2}\avg{\qka \qkkaa},\\
	\lkaka^{PQ} &=& \JBM\avg{\pka \qkkaa},\\
	\nka^P &=& \sqrt{\JOBM}\avg{\pka},\\
	\nka^Q &=& \sqrt{\JOBM}\avg{\qka},
\end{eqnarray}
\end{subequations}
in which the averages are taken with the effective Hamiltonian.
In the latter, the first two terms (i.e., those $\propto N^{-1}$) vanished in the thermodynamic limit.

Due to symmetries present in the system, we have
\begin{equation}
	\avg{\pka \pkkaa} = \avg{\qka \qkkaa}
	\quad\mbox{ and }\quad \avg{\pka} = \avg{\qka},
\end{equation}
which allows us to reduce the number of order parameters,
\begin{equation}
	\lkaka\equiv\lkaka^P = \lkaka^Q,\quad
	\nka\equiv \nka^P = \nka^Q.
\end{equation}
The expression for the effective Hamiltonian reads now
\begin{multline}
	-\beta\Heff = \JBM\Skaka\Lr[\Bigg]{
		\lkaka\lr*{\pka \pkkaa + \qka \qkkaa}\\
		+ \lkaka^{PQ}\pka \qkkaa
	}\\
	 +\sqrt{\JOBM}\Ska \nka\lr*{\pka + \qka} +\Mpq.
\end{multline}

\subsection{Decomposition into static and dynamic variables}
The variables $\lambda$ are of two types, according to the decomposition:
\begin{equation}\label{app:eq:decomp}
	\lkaka = \JBMa{2}\Lr[\Big]{\Rkk\delta_{\alpha \alpha'}
		+ \lr{1-\delta_{\alpha \alpha'}}\Qaa}.
\end{equation}
Terms with $\alpha=\alpha'$, denoted by $\Rkk$,
represent dynamic self-interactions that depend only on the difference $\abs{k-k'}$
due to the time-translational invariance,
while those with $\alpha\neq\alpha'$ ($\Qaa$) are
purely static and related to the EA order parameter,
\begin{equation}
	{\cal Q}_{\rm EA}=\lim_{n\to 0}\frac{2}{n(n-1)}\sum_{\alpha>\alpha'}\Qaa.
\end{equation}
We further simplify them to $\Qaa=q$ by assuming replica symmetry.
A similar decomposition can be applied to the cross-correlation term, i.e.,
\begin{equation}
	\lkaka^{PQ} = \JBM\Lr[\Big]{\Ukk\delta_{\alpha \alpha'}
		+ \lr{1-\delta_{\alpha \alpha'}}u}.
\end{equation}
In the same manner, the dependence of $\nka$ on its indices is dropped, i.e.,
\begin{equation}
	\nka = \sqrt{\JOBM}\varDelta.
\end{equation}
The equations now take the form
\begin{multline}
	\mathcal{F} = \frac{n}{2}\lr*{\JBM}^2 \sum_{kk'}(\Rkk^2+\Ukk^2)
		+ \frac{n(n-1)}{2} \lr*{J\beta}^{2} (q^2+u^2)\\
		+ n J_0\beta\varDelta^2
		- \ln \Trpq \eBH
\end{multline}
with
\begin{multline}
	-\beta \Heff = 
	\frac{1}{2}\lr*{\JBM}^{2}\sum_{kk'\alpha}\Lr[\Bigg]{
		\lr*{\Rkk-q}\lr*{ \pka\pkka + \qka\qkka }\\
		+2(\Ukk-u)\pka\qkka
	}\\
	+\frac{q}{2} \lr*{\JBM}^{2}\Lr*{\lr*{\Ska\pka}^2 + \lr*{\Ska\qka}^2}\\
	+u{\JBM}^{2}\lr*{\Ska\pka}\lr*{\Ska\qka}\\
	+\varDelta\JOBM\Ska \lr*{\pka+\qka}
	+\ln \mathcal{M}_{pq}.
\end{multline}

\subsection{Taking the $n \to 0$ limit}
To get rid of the replica-mixing terms,
we apply the Hubbard-Stratonovich transformation again, and get
\begin{multline}
	\mathcal{F} = \frac{n}{2}\lr*{\JBM}^2 \sum_{kk'}(\Rkk^2+\Ukk^2)\\
		+ \frac{n(n-1)}{2} \lr*{J\beta}^{2} (q^2+u^2)
		+ n J_0\beta\varDelta^2\\
		- \ln \Trpq
			\iiint\limits_{-\infty}^{\infty}\!\!\mathrm{d}x_P\mathrm{d}x_Q\mathrm{d}x_B
			\exp \lr*{ -\beta \mathcal{H}}\\
			\times \exp(-x_P^2-x_Q^2-x_B^2),
\end{multline}
where
\begin{multline}
	-\beta\mathcal{H} = \frac{1}{2}\lr*{\JBM}^{2}\sum_{kk'\alpha}\Lr[\Bigg]{
		\lr*{\Rkk-q}\lr*{ \pka\pkka + \qka\qkka }\\
		+2(\Ukk-u)\pka\qkka
	}\\
	+2\JBM\Lr[\Bigg]{
		\sqrt{\frac{u}{2}}x_{B}\lr*{\Ska\pka+\Ska\qka}\\
		+\sqrt{\frac{q-u}{2}}\lr*{x_{P}\Ska\pka+x_{Q}\Ska\qka}
	}\\
	+\varDelta\JOBM\Ska \lr*{\pka+\qka} +\ln \mathcal{M}_{pq}.
\end{multline}
Taking the limit of $n\rightarrow 0$ results in
\begin{multline}
	\mathcal{F} = \frac{1}{2}\lr*{\JBM}^2 \sum_{kk'}(\Rkk^2+\Ukk^2)
		- \frac{1}{2} \lr*{J\beta}^{2} (q^2+u^2)
		+ J_0\beta\varDelta^2\\
		- \DxDy \ln \Trpq \exp\lr*{-\beta \mathcal{H}},
\end{multline}
where we use the notation $\mathrm{D}x \equiv \mathrm{d}x\exp(-x^2)$.
The associated effective Hamiltonian reads
\begin{multline}
	-\beta\mathcal{H} =\frac{1}{2}\lr*{\JBM}^{2}\sum_{kk'}\Lr[\Big]{
		\lr*{\Rkk-q}\lr*{ p_{k}p_{k'} + q_{k}q_{k'} }\\
		+2(\Ukk-u)p_{k}q_{k'}
	}\\
	+2\JBM\Lr[\Bigg]{
		\sqrt{\frac{u}{2}}x_{B}\lr*{\sum_{k}p_{k}+\sum_{k}q_{k}}\\
		+\sqrt{\frac{q-u}{2}}\lr*{x_{P}\sum_{k}p_{k}+x_{Q}\sum_{k}q_{k}}
	}\\
	+ \varDelta\JOBM\sum_{k} \lr*{p_{k}+q_{k}} +\ln \mathcal{M}_{pq},
\end{multline}
while the final form of the self-consistent equations is
\begin{subequations}
\begin{eqnarray}
	\Rkk &=& \DxDy \avg{p_kp_{k'}},\\
	\Ukk &=& \DxDy \avg{p_kq_{k'}},\\
	\varDelta &=& \DxDy \avg{p_k},\\
	\label{app:eq:selfcon_q} q &=& \DxDy \avg{p_k}^2,\\
	\label{app:eq:selfcon_u} u &=& \DxDy \avg{p_k}\avg{q_k}.
\end{eqnarray}
\end{subequations}

\section{Self consistent equation for $q$}\label{app:selfconq}
For the following sections, we introduce the shorthands
\begin{equation}
	\SP \equiv \sum_{k}p_{k}, \qquad \SQ \equiv \sum_{k}q_{k}, \qquad \SJ \equiv \JBM
\end{equation}
as well as
\begin{equation}
    \DxDy \mathcal{A} \equiv \Dxxx \mathcal{A}.
\end{equation}
The penultimate equation \eqref{app:eq:selfcon_q} of the previous section comes from the following derivation:
\begin{widetext}
\begin{equation}\begin{split}\label{app:eq:dFdq}
	0 &= \frac{\partial F}{\partial q} = -\SJ^2M^2q - \Dxxx \Lr*{
		-\frac{\SJ^2}{2} \avg*{\SP\SP+\SQ\SQ}
		+\frac{\SJ}{2\sqrt{\frac{q-u}{2}}} \avg*{x_{P}\SP+x_{Q}\SQ}
	}\\
	&= -\SJ^2M^{2}q
		+\frac{\SJ^2}{2} \Dxxx \avg*{\SP\SP+\SQ\SQ}
		-\frac{\SJ}{2\sqrt{\frac{q-u}{2}}} \Dxxx \lr*{x_{P}\avg*{\SP}+x_{Q}\avg*{\SQ}}\\
	{}^{(*)} &= -\SJ^2M^{2}q
		+\frac{\SJ^2}{2} \Dxxx \avg*{\SP\SP+\SQ\SQ}
		-\frac{\SJ}{2\sqrt{\frac{q-u}{2}}} \Dxxx \SJ\sqrt{\frac{q-u}{2}}\lr*{\avg*{\SP\SP}-\avg*{\SP}\avg*{\SP}+\avg*{\SQ\SQ}-\avg*{\SQ}\avg*{\SQ}}\\
	&= -\SJ^2M^{2}q
		+\frac{\SJ^2}{2} \Dxxx \avg*{\SP\SP+\SQ\SQ}
		-\frac{\SJ^2}{2} \Dxxx \lr*{\avg*{\SP\SP}-\avg*{\SP}\avg*{\SP}+\avg*{\SQ\SQ}-\avg*{\SQ}\avg*{\SQ}}\\
	&= -\SJ^2M^{2}q
		+\frac{\SJ^2}{2} \Dxxx \lr[\Big]{\avg*{\SP}\avg*{\SP} + \avg*{\SQ}\avg*{\SQ}}
	= -\SJ^2M^{2}q
		+\frac{\SJ^2}{2} \sum_{kk'} \Dxxx \lr[\Big]{\avg*{p_k}\avg*{p_{k'}} + \avg*{q_k}\avg*{q_{k'}}},
\end{split}\end{equation}
where the transition marked with $(*)$ was done via integrating by parts.
The equation \eqref{app:eq:selfcon_u} is obtained in an analogous manner from $\frac{\partial F}{\partial u} = 0$.

\section{Glass critical line condition}\label{app:q}
We obtain the critical line conditions from the Landau theory condition that
\begin{equation}
	\frac{\partial^2 F}{\partial q^2}\Biggr|_{q=0} = 0
\end{equation}
at the critical temperature.
To expand the free energy \eqref{eq:FreeEnergy} to the second order in $q$,
we use the expression for the first derivative found in \eqref{app:eq:dFdq}
and differentiate it for the second time:
\begin{equation}\begin{split}\label{app:eq:d2Fdq2a}
	\frac{\partial^2 F}{\partial q^2} =& -\SJ^2M^{2}
		+ \frac{\SJ^2}{2} \Dxxx \lr[\Bigg]{2\avg*{\SP}\frac{\partial\avg{\SP}}{\partial q} + 2\avg*{\SQ}\frac{\partial\avg{\SQ}}{\partial q}}\\
	=& -\!\SJ^2M^2\!
		+ \SJ^2 \!\!\!\Dxxx \LR[\Bigg]{
		  \avg*{\SP}\avg*{\SP \,    \Lr[\Bigg]{\frac{\SJ}{2\sqrt{\frac{q-u}{2}}}(x_{P}\SP+x_{Q}\SQ) - \frac{\SJ^2}{2}(\SP\SP+\SQ\SQ)}}
		- \avg*{\SP}\avg*{\SP}\avg*{\frac{\SJ}{2\sqrt{\frac{q-u}{2}}}(x_{P}\SP+x_{Q}\SQ) - \frac{\SJ^2}{2}(\SP\SP+\SQ\SQ)}\\
		&+ \avg*{\SQ}\avg*{\SP\,    \Lr[\Bigg]{\frac{\SJ}{2\sqrt{\frac{q-u}{2}}}(x_{P}\SP+x_{Q}\SQ) - \frac{\SJ^2}{2}(\SP\SP+\SQ\SQ)}}
		- \avg*{\SQ}\avg*{\SP}\avg*{\frac{\SJ}{2\sqrt{\frac{q-u}{2}}}(x_{P}\SP+x_{Q}\SQ) - \frac{\SJ^2}{2}(\SP\SP+\SQ\SQ)}}\\
	=& -\SJ^2M^2
		- \frac{\SJ^4}{2} \Dxxx \avg*{\SP}\Lr*{\avg*{\SP(\SP\SP+\SQ\SQ)}-\avg*{\SP}\avg*{\SP\SP+\SQ\SQ}} 
		- \frac{\SJ^4}{2} \Dxxx \avg*{\SQ}\Lr*{\avg*{\SQ(\SP\SP+\SQ\SQ)}-\avg*{\SQ}\avg*{\SP\SP+\SQ\SQ}} \\
		&+ \frac{\SJ^3}{2\sqrt{\frac{q-u}{2}}} \Dxxx \, x_{P} \lr[\Bigg]{\avg*{\SP}\avg*{\SP\SP}-\avg*{\SP}\avg*{\SP}\avg*{\SP} + \avg*{\SQ}\avg*{\SQ\SP}-\avg*{\SQ}\avg*{\SQ}\avg*{\SP}} \\
		&+ \frac{\SJ^3}{2\sqrt{\frac{q-u}{2}}} \Dxxx \, x_{Q} \lr[\Bigg]{\avg*{\SP}\avg*{\SP\SQ}-\avg*{\SP}\avg*{\SP}\avg*{\SQ} + \avg*{\SQ}\avg*{\SQ\SQ}-\avg*{\SQ}\avg*{\SQ}\avg*{\SQ}}.
\end{split}\end{equation}
We focus on the last term of the above expression and integrate it by parts:
\begin{equation}\begin{split}
	\!\!\!\!\!\!\frac{\SJ^3}{2\sqrt{\frac{q-u}{2}}} \Dxxx \, x_{Q} & \lr[\Bigg]{\avg*{\SP}\avg*{\SP\SQ}-\avg*{\SP}\avg*{\SP}\avg*{\SQ} + \avg*{\SQ}\avg*{\SQ\SQ}-\avg*{\SQ}\avg*{\SQ}\avg*{\SQ}}\\
	=& \frac{\SJ^3}{4\sqrt{\frac{q-u}{2}}} \Dxxx \frac{\partial}{\partial x_{Q}} \lr[\Bigg]{\avg*{\SP}\avg*{\SP\SQ}-\avg*{\SP}\avg*{\SP}\avg*{\SQ} + \avg*{\SQ}\avg*{\SQ\SQ}-\avg*{\SQ}\avg*{\SQ}\avg*{\SQ}}\\
	=&\frac{\SJ^4}{2}\!\Dxxx \Lr[\Bigg]{
		  \lr[\Bigg]{ \avg*{\SP\SQ}\avg*{\SP\SQ}+\avg*{\SP}\avg*{\SP\SQ\SQ}-2\avg*{\SP}\avg*{\SP\SQ}\avg*{\SQ} }\\
		&\qquad\qquad - \lr[\Bigg]{ \avg*{\SP\SQ}\avg*{\SP}\avg*{\SQ}+\avg*{\SP}\avg*{\SP\SQ}\avg*{\SQ}+\avg*{\SP}\avg*{\SP}\avg*{\SQ\SQ}-3\avg*{\SP}\avg*{\SP}\avg*{\SQ}\avg*{\SQ} }\\
		&\qquad\qquad + \lr[\Bigg]{ \avg*{\SQ\SQ}\avg*{\SQ\SQ}+\avg*{\SQ}\avg*{\SQ\SQ\SQ}-2\avg*{\SQ}\avg*{\SQ\SQ}\avg*{\SQ} }\\
		&\qquad\qquad - \lr[\Bigg]{ \avg*{\SQ\SQ}\avg*{\SQ}\avg*{\SQ}+\avg*{\SQ}\avg*{\SQ\SQ}\avg*{\SQ}+\avg*{\SQ}\avg*{\SQ}\avg*{\SQ\SQ}-3\avg*{\SQ}\avg*{\SQ}\avg*{\SQ}\avg*{\SQ} }
	}\\
	=& \frac{\SJ^4}{2} \Dxxx \Lr[\Bigg]{
		  \lr[\Bigg]{ \avg*{\SP\SQ}\avg*{\SP\SQ}+\avg*{\SP}\avg*{\SP\SQ\SQ}-4\avg*{\SP}\avg*{\SP\SQ}\avg*{\SQ} 
		-  \avg*{\SP}\avg*{\SP}\avg*{\SQ\SQ}+3\avg*{\SP}\avg*{\SP}\avg*{\SQ}\avg*{\SQ} }\\
		&\qquad\qquad + \lr[\Bigg]{ \avg*{\SQ\SQ}\avg*{\SQ\SQ}+\avg*{\SQ}\avg*{\SQ\SQ\SQ}-5\avg*{\SQ}\avg*{\SQ\SQ}\avg*{\SQ} +3\avg*{\SQ}\avg*{\SQ}\avg*{\SQ}\avg*{\SQ} }
	}.
\end{split}\end{equation}
Plugging it back to \eqref{app:eq:d2Fdq2a} yields
\begin{equation}\begin{split}
	\frac{\partial^2 F}{\partial q^2} =& -\SJ^2M^2
		- \frac{\SJ^4}{2} \Dxxx \avg*{\SP}\Lr[\Bigg]{\avg*{\SP(\SP\SP+\SQ\SQ)}-\avg*{\SP}\avg*{\SP\SP+\SQ\SQ}} 
		- \frac{\SJ^4}{2} \Dxxx \avg*{\SQ}\Lr[\Bigg]{\avg*{\SQ(\SP\SP+\SQ\SQ)}-\avg*{\SQ}\avg*{\SP\SP+\SQ\SQ}} \\
		&+ \frac{\SJ^4}{2} \Dxxx \Lr[\Bigg]{
			\lr[\Big]{ \avg*{\SQ\SP}\avg*{\SQ\SP}+\avg*{\SQ}\avg*{\SQ\SP\SP}-4\avg*{\SQ}\avg*{\SQ\SP}\avg*{\SP} 
			-  \avg*{\SQ}\avg*{\SQ}\avg*{\SP\SP}+3\avg*{\SQ}\avg*{\SQ}\avg*{\SP}\avg*{\SP} }\\
			&+ \lr[\Big]{ \avg*{\SP\SP}\avg*{\SP\SP}+\avg*{\SP}\avg*{\SP\SP\SP}-5\avg*{\SP}\avg*{\SP\SP}\avg*{\SP} +3\avg*{\SP}\avg*{\SP}\avg*{\SP}\avg*{\SP} }
		}\\
		&+ \frac{\SJ^4}{2} \Dxxx \Lr[\Bigg]{
			\lr[\Big]{ \avg*{\SP\SQ}\avg*{\SP\SQ}+\avg*{\SP}\avg*{\SP\SQ\SQ}-4\avg*{\SP}\avg*{\SP\SQ}\avg*{\SQ} 
			-  \avg*{\SP}\avg*{\SP}\avg*{\SQ\SQ}+3\avg*{\SP}\avg*{\SP}\avg*{\SQ}\avg*{\SQ} }\\
			&+ \lr[\Big]{ \avg*{\SQ\SQ}\avg*{\SQ\SQ}+\avg*{\SQ}\avg*{\SQ\SQ\SQ}-5\avg*{\SQ}\avg*{\SQ\SQ}\avg*{\SQ} +3\avg*{\SQ}\avg*{\SQ}\avg*{\SQ}\avg*{\SQ} }
		}\\
	=& -\!\SJ^2M^2 + \frac{\SJ^4}{2} \!\Dxxx \lr[\Bigg]{
		\avg*{\SP\SP}^{\!2} \!\!- 4\avg*{\SP}^{\!2}\!\!\avg*{\SP\SP} + 3\avg*{\SP}^{\!4}\!\!
		+2\avg*{\SP\SQ}^{\!2} \!\!- 8\avg*{\SP}\avg*{\SQ}\avg*{\SP\SP} + 6\avg*{\SP}^{\!\!2}\!\!\avg*{\SQ}^{\!2}\!
		+\avg*{\SQ\SQ}^{\!2}\!\! - 4\avg*{\SQ}^{\!2}\!\!\avg*{\SQ\SQ} + 3\avg*{\SQ}^{\!4}
	}.
\end{split}\end{equation}
\end{widetext}
We obtain the critical line from the requirement
\begin{equation}
	\frac{\partial^2 F}{\partial q^2}\Biggr|_{q=0} = 0.
\end{equation}
The resulting equation reads
\begin{multline}
	1 = \frac{\SJ^2}{2M^2} \Dxxx \lr[\Bigg]{
		\avg*{\SP\SP}^2 - 4\avg*{\SP}^2\avg*{\SP\SP} + 3\avg*{\SP}^4\\
		+2\avg*{\SP\SQ}^2 - 8\avg*{\SP}\avg*{\SQ}\avg*{\SP\SP} + 6\avg*{\SP}^2\avg*{\SQ}^2\\
		+\avg*{\SQ\SQ}^2 - 4\avg*{\SQ}^2\avg*{\SQ\SQ} + 3\avg*{\SQ}^4
	}.
\end{multline}
Here, we also note that in the case of $J_0=0$ the condition reduces to
\begin{equation}
	\sum_{k}\Rkk = \frac{1}{2},
\end{equation}
which is the same expression as the one found for a simpler model \cite{Piekarska2018_PRL120}.

\section{Superfluid critical line condition}\label{app:Delta}
Similarily, we find the critical line for the superfluid transition
by taking the second derivative with respect to $\varDelta$,
\begin{multline}
	\frac{\partial F}{\partial \varDelta}
	= 2 J_{0}\beta\varDelta - \Dxxx \avg*{\JOBM\sum_{k}(p_{k}+q_{k})}\\
	= 2 J_{0}\beta\varDelta - \JOBM \Dxxx \avg*{\SP+\SQ}
\end{multline}
and
\begin{multline}
	0 = \frac{\partial^2 F}{\partial \varDelta^2} =
		2 J_{0}\beta - \lr*{\JOBM}^2 \Dxxx \Lr[\Bigg]{ \avg*{\SP(\SP+\SQ)}\\
		- \avg*{\SP}\avg*{\SP+\SQ} + \avg*{\SQ(\SP+\SQ)} - \avg*{\SQ}\avg*{\SP+\SQ} },
\end{multline}
which yields the critical-line condition
\begin{equation}
	1 = \frac{J_0\beta}{2M^2} \Dxxx \Lr*{ \avg*{(\SP+\SQ)(\SP+\SQ)} - \avg*{\SP+\SQ}^2}.
\end{equation}

\bibliography{all}
\bibliographystyle{apsrev4-2}
\end{document}